\documentclass[12pt]{article}

\usepackage{amsfonts,amssymb,bm,cite,amsmath}
\usepackage[dvips]{graphicx}
\usepackage[dvips]{psfrag}

\newcommand{\cg}[3]{C^{#1}_{#2\;#3}}

\newcommand {\beq}{\begin{equation}}
\newcommand {\eeq}{\end{equation}}
\newcommand {\beqa}{\begin{eqnarray}}
\newcommand {\eeqa}{\end{eqnarray}}
\newcommand {\n}{\nonumber \\}
\newcommand {\tr}{\mbox{tr}}

\newcommand{\sixj}[6]{{\scriptsize
			\begin{Bmatrix}
			#1 &\hspace{-2mm} #2 &\hspace{-2mm} #3 \\
			#4 &\hspace{-2mm} #5 &\hspace{-2mm} #6
			\end{Bmatrix}
			}}
\newcommand{\ninej}[9]{{\scriptsize
			\begin{Bmatrix}
			#1 &\hspace{-2mm} #2 &\hspace{-2mm} #3 \\
			#4 &\hspace{-2mm} #5 &\hspace{-2mm} #6 \\
			#7 &\hspace{-2mm} #8 &\hspace{-2mm} #9
			\end{Bmatrix}
			}}
			
\newcommand{\thalf}{{\tiny \mbox{$\frac{1}{2}$}}}

\newcommand{\cD}{{\cal D}}

\newcommand{\cF}{{\cal F}}
\newcommand{\cG}{{\cal G}}

\allowdisplaybreaks

\begin{document}
\setlength{\oddsidemargin}{0cm}
\setlength{\baselineskip}{7mm}

\begin{titlepage}
\renewcommand{\thefootnote}{\fnsymbol{footnote}}
\begin{normalsize}
\begin{flushright}
\begin{tabular}{l}
OU-HET 608\\
KEK-TH-1258\\
July 2008
\end{tabular}
\end{flushright}
  \end{normalsize}

~~\\

\vspace*{0cm}
    \begin{Large}
       \begin{center}
         {${\cal N}=4$ Super Yang-Mills
          from the Plane Wave Matrix Model}
       \end{center}
    \end{Large}
\vspace{1cm}

\begin{center}
Takaaki I{\sc shii}$^{1)}$\footnote
            {
e-mail address : 
ishii@het.phys.sci.osaka-u.ac.jp},
           Goro I{\sc shiki}$^{1,2)}$\footnote
            {
e-mail address : 
ishiki@post.kek.jp},
Shinji S{\sc himasaki}$^{1)}$\footnote
            {
e-mail address : 
shinji@het.phys.sci.osaka-u.ac.jp}
    {\sc and}
           Asato T{\sc suchiya}$^{3)}$\footnote
           {
e-mail address : satsuch@ipc.shizuoka.ac.jp}\\
      \vspace{1cm}
                    
       $^{1)}$ {\it Department of Physics, Graduate School of  
                     Science}\\
               {\it Osaka University, Toyonaka, Osaka 560-0043, Japan}\\
      \vspace{0.3cm}
       $^{2)}$ {\it Institute of Particle and Nuclear Studies}\\
               {\it High Energy Accelerator Research Organization (KEK)}\\
               {\it 1-1 Oho, Tsukuba, Ibaraki 305-0801, Japan}\\
      \vspace{0.3cm}
       $^{3)}$ {\it Department of Physics, Shizuoka University}\\
               {\it 836 Ohya, Suruga-ku, Shizuoka 422-8529, Japan}
               
\end{center}

\vspace{1cm}

\begin{abstract}
\noindent
We propose a nonperturbative definition of ${\cal N}=4$ super Yang-Mills (SYM).
We realize ${\cal N}=4$ SYM on $R\times S^3$ as the 
theory around a vacuum of the plane wave matrix model.
Our regularization preserves sixteen supersymmetries and the gauge symmetry.
We perform the 1-loop calculation to give evidences that the superconformal
symmetry is restored in the continuum limit.
\end{abstract}
\vfill
\end{titlepage}
\vfil\eject

\setcounter{footnote}{0}


\section{Introduction}
\setcounter{equation}{0}
\renewcommand{\thefootnote}{\arabic{footnote}} 
The AdS/CFT correspondence \cite{Maldacena,GKP,Witten}, a typical example of which is 
a conjecture that type IIB superstring on $AdS_5\times S^5$
corresponds to ${\cal N}=4$ super Yang-Mills (SYM), has been intensively investigated for a decade.
However, it has not been completely proven yet, partially because it is a strong/weak duality
with respect to the coupling constants. It is, therefore, relevant to give a nonperturbative
definition of ${\cal N}=4$ SYM that enables us to study its strong coupling regime.
The lattice gauge theory is a promising candidate for such a nonperturbative definition.
However, supersymmetric gauge theories on the lattice are generally difficult to construct, although
there have been remarkable developments on this subject 
\cite{Kaplan,Itoh,Catterall,Sugino,D'Adda}.
To give a nonperturbative definition of ${\cal N}=4$ SYM will not only bring
enormous progress in the study of the AdS/CFT correspondence, but will also
yield some insights into the problem of nonperturbative formulation 
of supersymmetric gauge theories.

It was shown in \cite{EK} that a gauge theory in the planar limit is equivalent to the matrix model (the
reduced model) obtained 
by dimensionally reducing it to zero dimension if the $U(1)^D$ symmetry is unbroken, where $D$ stands for
the dimensionality of space-time. This is the so-called large $N$ reduction.
The global gauge symmetry of the matrix model is naturally interpreted as the local gauge symmetry of the 
original gauge theory. Thus, as an alternative to the lattice gauge theory, the matrix model may serve as 
a nonperturbative definition of the planar gauge theory with the gauge symmetry manifestly kept.
The $U(1)^D$ symmetry is, however, spontaneously broken except for $D=2$, so that the above equivalence
does not hold generically. There have been two improvements of the reduced model in which
the $U(1)^D$ symmetry breaking is prevented so that the equivalence holds: 
one is the quenched reduced model \cite{Bhanot,Parisi,Gross,Das}, and 
the other is the twisted reduced model \cite{GonzalezArroyo}.
These improved models work well for nonsupersymmetric 
planar gauge theories\footnote{Recent studies on the twisted 
\cite{Teper, Azeyanagi,Bietenholz:2007xh} and
quenched \cite{Bringoltz} reduced models
of the lattice gauge theory
oppose this statement, and an improvement of the reduced model was
studied in \cite{Unsal}. Anyway, these studies do not affect the arguments
in this paper directly, 
because we consider a different kind of 
reduced model 
and our model has supersymmetry.}. It seems quite difficult to preserve
supersymmetry manifestly in the twisted reduced model on the flat space and
in the quenched reduced model, while the gauge symmetry is respected
in both models.

The compactification in matrix models developed in \cite{Taylor} shares the same idea with the reduced model and
will be called the matrix T-duality in this paper.
While it is not restricted to the planar limit, it requires the size of matrices to be infinite from the 
beginning for the 
orbifolding condition to be imposed, so that it cannot be used to define any supersymmetric 
gauge theory nonperturbatively as it stands.
It was argued in \cite{Kaplan} that by imposing an orbifolding condition on the reduced model of 
a supersymmetric gauge theory, one can obtain its lattice theory in which part of the supersymmetries
are manifestly preserved so that the fine-tuning of only a few parameters is required. 
This construction can be regarded as
a finite-size matrix analog of the matrix T-duality. However, it has a problem of flat directions which is analogous to the problem of the $U(1)^D$ symmetry breaking. 
To overcome this problem, for instance, one needs to introduce a mass term
for the scalar field, which leads to no preservation of supersymmetries. 

In \cite{ISTT}, Takayama and three of the present authors found
the relationships among the $SU(2|4)$ symmetric theories, which
include ${\cal N}=4$ SYM on $R\times S^3/Z_k$, 
2+1 SYM on $R\times S^2$ \cite{Maldacena:2002rb}
and the plane wave matrix model (PWMM) \cite{Berenstein:2002jq}. 
The last theory is  obtained by consistently truncating the Kaluza-Klein modes of
${\cal N}=4$ SYM on $R\times S^3$ \cite{Kim-Klose-Plefka} and so are the former two theories \cite{Lin-Maldacena}. 
In particular, 2+1 SYM on $R\times S^2$ and PWMM can be regarded as 
dimensional reductions of ${\cal N}=4$ SYM on $R\times S^3$.
These theories possess common features: mass gap,
discrete spectrum and many discrete vacua. From the gravity duals of those vacua proposed
by Lin and Maldacena \cite{Lin-Maldacena},
the following relations among these theories are suggested: 
A) the theory around each vacuum
of 2+1 SYM on $R\times S^2$ is equivalent to the theory around a certain
vacuum of PWMM, and B) the theory around each vacuum
of ${\cal N}=4$ SYM on $R\times S^3/Z_k$ is equivalent to the theory around
a certain vacuum of 2+1 SYM on $R\times S^2$ with the orbifolding
(periodicity) condition imposed.
In \cite{ISTT}, the relations A) and B) were shown directly on the gauge theory side.
The results in \cite{ISTT} not only serve as a nontrivial check of the gauge/gravity 
correspondence for the $SU(2|4)$ theories, but they are also interesting from the point of view of the reduced model
as follows. 
While there have been many works on realizing 
the gauge theories on the fuzzy sphere \cite{Madore:1991bw,Grosse:1992bm,Grosse:1995ar,CarowWatamura:1998jn} 
using matrix models \cite{Iso:2001mg,Steinacker:2003sd} and on the monopoles on the fuzzy sphere 
\cite{Steinacker:2003sd,Grosse:1995jt,Baez:1998he,Landi,CarowWatamura:2004ct,Aoki:2003ye},
the relation A) shows that
the continuum limit of the concentric fuzzy spheres with different radii corresponds to multiple monopoles.
Note that realizing the gauge theories on the fuzzy sphere using the matrix models can be viewed as an extension of the twisted reduced model 
to curved space.
The relation B) can be regarded as an extension of 
the matrix T-duality to that on a nontrivial $U(1)$ bundle, $S^3/Z_k$, whose base space is $S^2$.
Indeed, the matrix T-duality was later extended to that on general $U(1)$ bundles in \cite{IIST} and on general $SU(2)$ bundles in \cite{IIST2}.
Combining the relations A) and B) leads to the relation C), that the theory around each vacuum
of ${\cal N}=4$ SYM on $R\times S^3/Z_k$ is equivalent to the theory around a certain vacuum of PWMM
with the orbifolding condition imposed.
In particular, for $k=1$, ${\cal N}=4$ SYM on $R\times S^3$ 
is realized in PWMM.
The possibility of defining ${\cal N}=4$ SYM in terms of PWMM nonperturbatively
is suggested in \cite{ISTT}.
The relationships shown in \cite{ISTT} are classical in the following sense: in the relation A), we show the equivalence
at tree level and do not care about possible UV/IR mixing at higher orders, although the gravity duals suggest
that any UV/IR mixing does not occur. In the relation B), the size of
matrices must be infinite from the beginning as in the original matrix T-duality.

In this paper, we propose a nonperturbative definition of ${\cal N}=4$ SYM on $R\times S^3$ which is 
equivalently mapped to ${\cal N}=4$ SYM on $R^4$ at the conformal point and possesses the superconformal
symmetry, the $SU(2,2|4)$ symmetry. We restrict ourselves to the planar limit.
By referring to the relation C) in \cite{ISTT},
we regularize ${\cal N}=4$ SYM on $R\times S^3$ nonperturbatively by using PWMM. Our analysis in this paper
is quantum mechanical.
The restriction to the planar limit enables us not to impose the orbifolding condition and to consider finite-size
matrices such that the size of matrices plays the role of the ultraviolet cutoff.
Thus we use an extension of the reduced model to curved space rather than the matrix T-duality
to relate $2+1$ SYM on $R\times S^2$ to ${\cal N}=4$ SYM on $R\times S^3$. 
Because PWMM is a massive theory, there is no flat direction and the quenching prescription is not needed. 
Our regularization manifestly preserves 
the gauge symmetry and the $SU(2|4)$ symmetry, a subgroup 
of the $SU(2,2|4)$ symmetry. 
In particular, sixteen supersymmetries among thirty-two supersymmetries
are respected in our regularization.
The restriction to the planar limit and
sixteen supersymmetries are probably sufficient to suppress
the UV/IR mixing which may break the relation between 
$2+1$ SYM on $R\times S^2$ and PWMM quantum mechanically.
They also stabilize the vacua of PWMM completely.
Indeed, the gravity duals of these theories suggest 
$2+1$ SYM on $R\times S^2$ is obtained from PWMM
quantum mechanically
in the continuum limit at least in the planar limit.
The full $SU(2,2|4)$ symmetry should be restored in the continuum limit.
By performing the 1-loop analysis and comparing the results with
those in continuum ${\cal N}=4$ SYM, we provide some evidences that
our regularization of ${\cal N}=4$ SYM
indeed works, although our final goal is to analyze ${\cal N}=4$ SYM 
nonperturbatively by using
our formulation. 
Our theory still has the continuum time direction, which we need to cope with 
in order to put our theory on computer.
For instance, we should be able to apply the method in \cite{Hanada,Anagnostopoulos,Catterall:2008yz} to our case.
We comment on an interesting paper \cite{Kaneko}, the authors of which constructed 
the $S^3$ background in the IIB matrix model with the Myers term  using the same procedure
as \cite{ISTT}. They calculated the free energy of the theory around the background up to the 2-loop order to
find the stability of the background. 
Note also that the authors of \cite{Elliott:2008jp} discussed practicality
of ${\cal N}=4$ SYM on the lattice recently.

This paper is organized as follows. 
In section 2, we study the large $N$ reduction on a finite volume. As an example, 
we consider the $\phi^4$ matrix quantum
mechanics.  We examine how the theory on $S^1$ is obtained from the matrix model that is its
dimensional reduction to zero dimension, emphasizing the difference between the large $N$ reductions
for the theories on $R$ and $S^1$. In section 3, we review the relationships among
${\cal N}=4$ SYM on $R\times S^3$, 2+1 SYM on $R\times S^2$ and PWMM shown
in \cite{ISTT}.
Based on these relationships and the result in section 2, we give a nonperturbative definition of 
${\cal N}=4$ SYM on $R\times S^3$ using PWMM. In section 4, we perform the 1-loop calculation in our 
theory to give some evidences that our regularization of ${\cal N}=4$ SYM indeed works.
Section 5 is devoted to conclusion and discussion. In appendices, some details are gathered.

\section{The large $N$ reduction on finite volume}
\setcounter{equation}{0}
In this section, we study the large $N$ reduction on a finite volume, focusing
on how different it is from that on an infinite volume. 
Let us consider a matrix quantum mechanics, whose
action is given by
\begin{align}
S=\int \:d\tau \mbox{Tr}\left(\frac{1}{2}\left(\frac{d\phi}{d\tau}\right)^2
+\frac{1}{2}m^2 \phi^2 +\frac{1}{4}g^2\phi^4 \right),
\label{original model}
\end{align}
where $\phi(\tau)$ is an $N\times N$ hermitian matrix.
We take the 't Hooft limit: $N\rightarrow \infty, \;\; \lambda=g^2N=\mbox{fixed}$.
First, we consider the case in which the theory is defined on $R$, 
namely $-\infty < \tau <\infty$.
The prescription of the large $N$ reduction is to make the
following replacement \cite{Parisi,Gross,Das}:
\begin{align}
&\phi(\tau) \rightarrow e^{iP\tau}\phi e^{-iP\tau}, \nonumber\\
&\int d\tau \rightarrow \frac{2\pi}{\Lambda},
\label{replacement}
\end{align}
where $\phi$ in the right-hand side of the first equation is no longer dependent on $\tau$
and $\Lambda$ is an ultraviolet cutoff.
$P$ is a constant $N\times N$ matrix given by
\begin{align}
P=\mbox{diag}(p_1,p_2,\cdots,p_N)
\label{P}
\end{align}
with $p_i=\frac{\Lambda}{N}(i-\frac{N}{2})$. 
We take the limit in which
\begin{align}
\Lambda \rightarrow \infty, \;\; N \rightarrow \infty, \;\; \Lambda/N \rightarrow 0.
\label{limit for reduced model}
\end{align}
Note that $\Lambda/N$ is an infrared cutoff.
The action (\ref{original model}) is reduced to 
\begin{align}
S_R=\frac{2\pi}{\Lambda}\left(\frac{1}{2}\sum_{i,j}((p_i-p_j)^2+m^2)|\phi_{ij}|^2+\frac{1}{4}g^2\mbox{Tr}(\phi^4)
\right).
\label{reduced model}
\end{align}

\begin{figure}[tbp]
\begin{center}
\begin{minipage}{0.2\hsize}
  \begin{center}
   {\includegraphics[width=60mm]{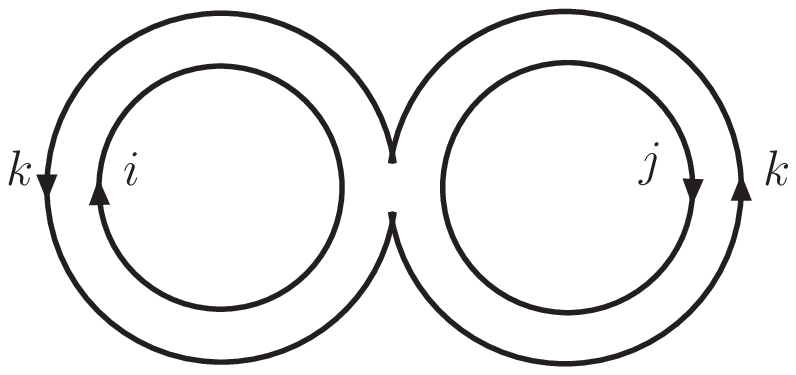}}
   (a)
  \label{G-aFig}
  \end{center}
 \end{minipage}
  \hspace*{3cm}
\begin{minipage}{0.2\hsize}
 \begin{center}
  {\includegraphics[width=50mm]{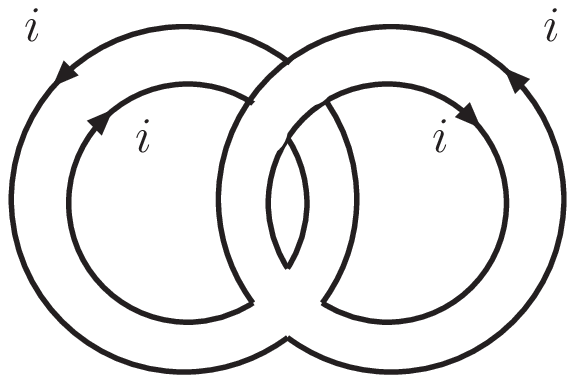}}
  (b)
  \label{G-bFig}
 \end{center}
\end{minipage}
\end{center}
\caption{2-loop diagrams for the free energy: (a)planar and (b)nonplanar}
\end{figure}

In order to illustrate the large $N$ reduction, we see that the free energy 
of the original model (\ref{original model}) agrees with that of the reduced model (\ref{reduced model}) at
the two-loop level. 
There are two diagrams at the two-loop level for the free energy. Fig. 1-(a) shows the planar diagram while Fig. 1-(b) shows the nonplanar one.
We evaluate the planar diagram in Fig. 1-(a) for the original model:
\begin{align}
F^{2-loop}_{planar}/\mbox{Vol}=
\frac{1}{2}N^2  \lambda \int \frac{dpdq}{(2\pi)^2} 
\frac{1}{(p^2+m^2)(q^2+m^2)}.
\label{planar diagram for the original model}
\end{align}
The nonplanar diagram in Fig. 1-(b)
for the original model is suppressed by the order of $1/N^2$ compared to the planar one in Fig. 1-(a).
On the other hand, we evaluate the planar diagram in Fig. 1-(a) for the reduced model:
\begin{align}
F^{2-loop}_{R,planar}/(2\pi/\Lambda)
&=\frac{1}{2}g^2 \left(\frac{\Lambda}{2\pi}\right)^2\sum_{i,j,k}\frac{1}{((p_i-p_k)^2+m^2)((p_j-p_k)^2+m^2)}
\nonumber\\
&=\frac{1}{2}g^2\left(\frac{\Lambda}{2\pi}\right)^2N
\sum_{i,j}\frac{1}{(p_i^2+m^2)(p_j^2+m^2)}.
\label{planar diagram for the reduced model}
\end{align}
By using the relation valid in the limit (\ref{limit for reduced model}),
\begin{align}
\frac{\Lambda}{N}\sum_if\left(\frac{\Lambda}{N}\left(i-\frac{N}{2}\right)\right)=\int_{-\infty}^{\infty} dp f(p),
\end{align}
one can easily verify that $F^{2-loop}_{planar}/\mbox{Vol}=F^{2-loop}_{R,planar}/(2\pi/\Lambda)$.
Indeed, one can prove that $F_{planar}/\mbox{Vol}=F_{R,planar}/(2\pi/\Lambda)$ holds at all orders.
We further evaluate the nonplanar diagram in Fig. 1-(b) for the reduced model:
\begin{align}
F^{2-loop}_{R,nonplanar}/(2\pi/\Lambda)&=\frac{1}{4}g^2\left(\frac{\Lambda}{2\pi}\right)^2 
\sum_i \frac{1}{m^4} \nonumber\\
&=\frac{1}{4}N^2 \lambda \frac{1}{m^4} \times \left(\frac{\Lambda}{2\pi N}\right)^2.
\label{nonplanar diagram for the reduced model}
\end{align}
Note that there is no correspondence for the nonplanar diagram between the original and reduced models.
The nonplanar contribution (\ref{nonplanar diagram for the reduced model}) is suppressed
by the factor $(\Lambda/N)^2$ in the limit
(\ref{limit for reduced model}), relative to the planar contribution (\ref{planar diagram for the reduced model}).
All the nonplanar contributions are indeed suppressed relative to the planar contributions in the reduced model.
The reduced model therefore reproduces the 't Hooft (planar) limit of the original model.

Next, we compactify the $\tau$-direction to $S^1$ with the radius $R$.
We evaluate the planar diagram in Fig. 1-(a) for the original model: 
\begin{align}
\tilde{F}^{2-loop}_{planar}=\frac{1}{2}N^2\tilde{\lambda}\sum_{n,l}
\frac{1}{(\frac{n^2}{R^2}+m^2)(\frac{l^2}{R^2}+m^2)},
\end{align}
where we take the $N\rightarrow\infty$ limit with $\tilde{\lambda}=g^2N/(2\pi R)$ fixed.
Note that the nonplanar diagram in Fig. 1-(b) for the original model is still suppressed by $1/N^2$
relative to the planar diagram in Fig. 1-(a).
Correspondingly, we consider the reduced model
\begin{align}
\tilde{S}_R
=\frac{1}{2}\sum_{i,j}((p_i-p_j)^2+m^2)|\phi_{ij}|^2+\frac{1}{4}g_R^2\mbox{Tr}(\phi^4)
\label{reduced model for S^1}
\end{align}
with $p_i=\frac{1}{R}\left(i-\frac{N}{2}\right)$. This naive reduced model turns out not to reproduce the 
original model on $S^1$.
The contribution of the planar diagram in Fig. 1-(a) to the free energy for this reduced model is
\begin{align}
\tilde{F}^{2-loop}_{R,planar}&=\frac{1}{2}g_R^2 \sum_{i,j,k}\frac{1}{(\frac{(i-k)^2}{R^2}+m^2)(\frac{(j-k)^2}{R^2}+m^2)}
\nonumber\\
&=\frac{1}{2}g_R^2N\sum_{n,l}\frac{1}{(\frac{n^2}{R^2}+m^2)(\frac{l^2}{R^2}+m^2)},
\label{planar diagram for the reduced model for S^1}
\end{align}
while that of the nonplanar diagram in Fig. 1-(b) is
\begin{align}
\tilde{F}^{2-loop}_{R,nonplanar}=\frac{1}{4}g_R^2\sum_i \frac{1}{m^4}=\frac{1}{4}g_R^2 N \frac{1}{m^4}.
\label{nonplanar diagram for the reduced model for S^1}
\end{align}
(\ref{nonplanar diagram for the reduced model for S^1}) is not suppressed relative to
(\ref{planar diagram for the reduced model for S^1}), because the infrared cutoff $1/R$ is finite in this case.
Thus the correspondence between the original
and reduced models fails in this case.

In the following, we modify the reduced model (\ref{reduced model}) to recover the correspondence.
The action of the modified model takes the same form as (\ref{reduced model for S^1}) 
while $\phi$ is a $N(T+1)\times N(T+1)$ matrix, $i,j$ run from 1 to $N(T+1)$ and $p_i$ is given
by the $i$-component of the matrix 
\begin{align}
\tilde{P}=\frac{1}{R}\mbox{diag}\left(-\frac{T}{2},-\frac{T}{2}+1,\cdots, \frac{T}{2}\right)\otimes 1_N.
\end{align}
Here $T$ is a positive even integer.
We take the limit in which $T\rightarrow \infty,\;\;N\rightarrow \infty,\;\; \tilde{\lambda}=g_R^2N=\mbox{fixed}$.
$T$ turns out to play the role of the ultraviolet cutoff for the momentum.
In the modified model, the contribution of the planar diagram in Fig. 1-(a) to the free energy is
\begin{align}
\tilde{F}^{2-loop}_{MR,planar}&=\frac{1}{2}g_R^2N^3 
\sum_{a,b,c=-\frac{T}{2}}^{\frac{T}{2}}\frac{1}{(\frac{(a-c)^2}{R^2}+m^2)(\frac{(b-c)^2}{R^2}+m^2)}
\nonumber\\
&=\frac{1}{2}N^2(T+1)\tilde{\lambda}\sum_{n,l}\frac{1}{(\frac{n^2}{R^2}+m^2)(\frac{l^2}{R^2}+m^2)}
\label{planar diagram for modified model}
\end{align}
Then, we see that 
$\tilde{F}^{2-loop}_{planar}=\tilde{F}^{2-loop}_{MR,planar}/(T+1)$.
Indeed, it is easily verified that $\tilde{F}_{planar}=\tilde{F}_{MR,planar}/(T+1)$ holds at all orders.
On the other hand, the contribution of the nonplanar diagram in Fig. 1-(b) to the free energy for the modified 
model is
\begin{align}
\tilde{F}^{2-loop}_{MR,nonplanar}=\frac{1}{4}g_R^2N \sum_{a=-\frac{T}{2}}^{\frac{T}{2}} \frac{1}{m^4}
=\frac{1}{4}(T+1)\tilde{\lambda}\frac{1}{m^4}.
\end{align}
This is suppressed by $1/N^2$ relative to (\ref{planar diagram for modified model}).
All the nonplanar contributions are indeed suppressed relative to the planar contributions in the modified model.
Hence, the modified model reproduces the 't Hooft (planar) limit of the original model on $S^1$.

For $D$-dimensional
pure Yang-Mills (YM),
the reduction analogous to (\ref{replacement}) leads to 
\begin{align}
&S_{YM}=\frac{1}{g^2}\int d^Dx \frac{1}{4}\mbox{Tr}[\partial_{\mu}-iA_{\mu},\partial_{\nu}-iA_{\nu}]^2 \n
&\rightarrow S_{YM,R}=-\frac{2\pi}{\Lambda}\frac{1}{4g^2}\mbox{Tr}[P_{\mu}+A_{\mu},P_{\nu}+A_{\nu}]^2,
\label{reduced model of YM}
\end{align}
where $P_{\mu}$ is the $D$-dimensional analogue of (\ref{P}).
It is known that the reduced model (\ref{reduced model of YM}) does not reproduce the original YM, because
the diagonal elements of $A_{\mu}$ are zero-dimensional massless fields
and instable enough to absorb $P_{\mu}$.
This is interpreted as the counterpart of 
the $U(1)^D$ symmetry breaking in the reduced model of the lattice gauge theory.
Usually, in order to overcome this problem, the eigenvalues of $P_{\mu}+A_{\mu}$ in (\ref{reduced model of YM}) are
fixed to $P_{\mu}$ \cite{Gross}. This is a quenching prescription. While the gauge symmetry is
respected in this prescription, supersymmetry is not. In the case we are concerned with in this paper,
we want to respect both symmetries simultaneously. 
We see how this problem is overcome in the next section.

\section{Realization of ${\cal N}=4$ SYM on $R\times S^3$ in terms of PWMM}
\setcounter{equation}{0}
\renewcommand{\thefootnote}{\arabic{footnote}} 
In this section, we review the relationships among ${\cal N}=4$ SYM on $R\times S^3$,
2+1 SYM on $R\times S^2$ and PWMM shown in \cite{ISTT}, and we  propose a nonperturbative 
definition of ${\cal N}=4$ SYM on $R\times S^3$, based on these relationships and the result in the
previous section.

\subsection{${\cal N}=4$ SYM on $R\times S^3$ and the $SU(2|4)$ theories}
The action of ${\cal N}=4$ SYM on $R\times S^3$ takes the form\footnote{In this paper, we change the notation 
used in \cite{ISTT,ITT} as follows: $Y_i\rightarrow X_i,\;\;X_{AB}\rightarrow \Phi_{AB},\;\;\phi \rightarrow \chi$.} 
\begin{align}
&S_{R\times S^3}=\frac{1}{g_{R\times S^3}^2}\int dt\frac{d\Omega_3}{(\mu/2)^3}
\: \mbox{Tr}\left(
-\frac{1}{4}F_{ab}F^{ab}-\frac{1}{2}D_a\Phi_{AB}D^a\Phi^{AB}
-\frac{1}{2}\Phi_{AB}\Phi^{AB}  \right. \n
&\hspace{3.7cm}+\frac{1}{4}[\Phi_{AB},\Phi_{CD}][\Phi^{AB},\Phi^{CD}]+i\psi_A^{\dagger}D_t\psi^A
+i\psi_A^{\dagger}\sigma^iD_i\psi^A \n 
&\hspace{3.7cm}\left.+\psi_A^{\dagger}\sigma^2[\Phi^{AB},(\psi_B^{\dagger})^T]
-\psi^{AT}\sigma^2[\Phi_{AB},\psi^B] \right). 
\label{actiontwo-component}
\end{align}
Here $a,b$ are the local Lorentz indices and run from 0 to 3. ``0" corresponds to the time, $t$.
$A,B$ are indices of the fundamental representation of $SU(4)$ and run from 1 to 4. $\Phi_{AB}=-\Phi_{BA}$ and 
$\Phi^{AB}=\frac{1}{2}\epsilon^{ABCD}\Phi_{CD}$ .
The radius of $S^3$ is $2/\mu$. This theory possesses the superconformal symmetry, the $SU(2,2|4)$ symmetry.
The action of 2+1 SYM on $R\times S^2$ takes the form
\begin{align}
 &S_{R\times S^2}
=\frac{1}{g_{R\times S^2}^2}\int dt\frac{d\Omega_2}{\mu^2} \mbox{Tr}\biggl(
\frac{1}{2}(D_t\vec{X}-i\mu\vec{L}^{(0)}A_t)^2
 -\frac{1}{2}(\mu \vec{X}+i(\mu \vec{L}^{(0)}\times \vec{X}-\vec{X}\times \vec{X}))^2 \nonumber \\
 &\hspace{2.5cm}
 +\frac{1}{2}D_t\Phi_{AB}D_t\Phi^{AB}
 +\frac{1}{2}\vec{{\cal D}}\Phi_{AB}\cdot\vec{{\cal D}}\Phi^{AB}
 -\frac{\mu^2}{8}\Phi_{AB}\Phi^{AB}+\frac{1}{4}[\Phi_{AB},\Phi_{CD}][\Phi^{AB},\Phi^{CD}]
  \nonumber \\
 &\hspace{2.5cm}
 +i\psi_{A}^\dagger D_t\psi^A
 -\psi_A^\dagger\vec{\sigma}\cdot \vec{\cal D}\psi^A
 -\frac{3\mu}{4}\psi_A^\dagger\psi^A
 +\psi_A^\dagger\sigma^2[\Phi^{AB},(\psi_B^\dagger)^T]
 -\psi^{AT}\sigma^2[\Phi_{AB},\psi^B]\biggr),
\label{S^2SU(4)form}
\end{align}
where 
\begin{align}
\vec{L}^{(0)}=-i\vec{e}_{\varphi}\partial_{\theta}+i\frac{1}{\sin\theta}\vec{e}_{\theta}\partial_{\varphi}
\label{angular momentum operator}
\end{align}
with $\vec{e}_r=(\sin\theta\cos\varphi,\sin\theta\sin\varphi,\cos\theta)$,
$\vec{e}_{\theta}=\frac{\partial \vec{e}_r}{\partial \theta}$ and 
$\vec{e}_{\varphi}=\frac{1}{\sin\theta}\frac{\partial \vec{e}_r}{\partial \varphi}$,
$\vec{{\cal D}}=\mu \vec{L}^{(0)}-[\vec{X},\;]$ and the radius of $S^2$ is $1/\mu$.
The action of PWMM takes the form
\begin{align}
S_{PW}
&=\frac{1}{g_{PW}^2}\int \frac{dt}{\mu^2}\: \mbox{Tr}
\left(
\frac{1}{2}(D_t X_i)^2-\frac{1}{2}(\mu X_i-\frac{i}{2}\epsilon_{ijk}[X_j,X_k])^2
+\frac{1}{2}D_t \Phi_{AB}D_t \Phi^{AB}-\frac{\mu^2}{8}\Phi_{AB}\Phi^{AB}
\right. \nonumber\\
&\qquad\qquad
+\frac{1}{2}[X_i,\Phi_{AB}][X_i,\Phi^{AB}]
+\frac{1}{4}[\Phi_{AB},\Phi_{CD}][\Phi^{AB},\Phi^{CD}]
+i\psi_A^\dagger D_t\psi^A \nonumber\\
&\qquad\qquad\left.
-\frac{3\mu}{4}\psi_A^\dagger\psi^A 
+\psi_A^\dagger \sigma^i[X_i,\psi^A]
+\psi_A^\dagger \sigma^2[\Phi^{AB},({\psi}_B^\dagger)^T]
-(\psi^A)^T \sigma^2 [\Phi_{AB},\psi^B]
\right).
\label{action_of_PWMM}
\end{align}
Both 2+1 SYM on $R\times S^2$ and PWMM possess the $SU(2|4)$ symmetry, which is a subgroup of the $SU(2,2|4)$ symmetry 
and has sixteen supercharges.

In the reminder of this section, for simplicity, we ignore the time component of the 
gauge field $A_t$ and the matter degrees of freedom, $\Phi_{AB}$ and $\psi^A$.
It is easy to include these degrees of freedom in the arguments. All the statements in the following are also
valid with these degrees of freedom.

\subsection{$S^3$ and $S^2$}
First, we summarize some useful facts about $S^3$ and $S^2$ (see also \cite{IIST2}).
We regard $S^3$ as the $SU(2)$ group manifold. We parameterize an element
of $SU(2)$ in terms of the Euler angles as
\begin{equation}
g=e^{-i\varphi \sigma_3/2}e^{-i\theta \sigma_2/2}e^{-i\psi \sigma_3/2},
\label{Euler angles}
\end{equation}
where $0\leq \theta\leq \pi$, $0\leq \varphi < 2\pi$, $0\leq \psi < 4\pi$.
The periodicity with respect to these angle variables is expressed as
\begin{align}
(\theta,\varphi,\psi)\sim (\theta,\varphi+2\pi,\psi+2\pi)\sim (\theta,\varphi,\psi+4\pi).
\end{align}
The isometry of $S^3$ is $SO(4)=SU(2)\times SU(2)$, and these two
$SU(2)$'s act on $g$ from left and right, respectively. 
Note that the superconformal group $SU(2,2|4)$ includes the $SO(4)$ group as a subgroup.
We construct the
right-invariant 1-forms, 
\begin{equation}
dgg^{-1}=-i\mu E^i \sigma_i/2,
\end{equation}
where the radius of $S^3$ is $2/\mu$. They are explicitly
given by 
\begin{eqnarray}
&&E^1=\frac{1}{\mu}(-\sin \varphi d\theta + \sin\theta\cos\varphi d\psi),\nonumber\\
&&E^2=\frac{1}{\mu}(\cos \varphi d\theta + \sin\theta\sin\varphi
 d\psi),\nonumber\\
&&E^3=\frac{1}{\mu}(d\varphi + \cos\theta d\psi),
\end{eqnarray}
and satisfy the Maurer-Cartan equation
\begin{equation}
dE^i-\frac{\mu}{2}\epsilon_{ijk}E^j\wedge E^k=0.\label{Maure-Cartan}
\end{equation}
The metric is constructed from $E^i$ as
\begin{equation}
ds^2=E^iE^i=\frac{1}{\mu^2}\left(
d\theta^2+\sin^2\theta d\varphi^2 +(d\psi+\cos\theta d\varphi)^2\right).
\label{metric of S^3}
\end{equation}
The Killing vectors dual to $E^i$ are given by
\begin{equation}
{\cal{L}}_i=-\frac{i}{\mu}E^M_i\partial_M,
\end{equation}
where $M=\theta,\varphi,\psi$ and $E^M_i$ are inverse of $E^i_M$. The explicit form of the Killing
vectors are
\begin{eqnarray}
&&{\cal{L}}_1=-i\left(-\sin\varphi\partial_{\theta}-\cot\theta\cos\varphi\partial_{\varphi}+\frac{\cos\varphi}{\sin\theta}\partial_{\psi}\right),\nonumber\\
&&{\cal{L}}_2=-i\left(\cos\varphi\partial_{\theta}-\cot\theta\sin\varphi\partial_{\varphi}+\frac{\sin\varphi}{\sin\theta}\partial_{\psi}\right),\nonumber\\
&&{\cal{L}}_3=-i\partial_{\varphi}.\label{Killing vector}
\end{eqnarray}
Because of the Maurer-Cartan equation (\ref{Maure-Cartan}), the Killing vectors satisfy the SU(2) algebra, $[{\cal{L}}_i,{\cal{L}}_j]=i\epsilon_{ijk}{\cal{L}}_k$.

One can also regard $S^3$ as a $U(1)$ bundle over $S^2=SU(2)/U(1)$.
$S^2$ is parametrized by $\theta$ and $\varphi$ and covered with two local patches:
the patch I defined by $0\leq\theta <\pi$ and the patch II defined by $0<\theta\leq\pi$.
In the following expressions, the upper sign is taken in the patch I while the lower sign in the patch II.
The element of $SU(2)$ in (\ref{Euler angles}) is decomposed as 
\begin{align}
&g=L\cdot h \nonumber\\
&\mbox{with}\;\;
L=e^{-i\varphi \sigma_3/2}e^{-i\theta \sigma_2/2}e^{\pm i\varphi \sigma_3/2}\;\;\mbox{and}\;\;
h=e^{-i(\psi\pm\varphi) \sigma_3/2} 
\end{align}
$L$ represents an element of $S^2$, while $h$ represents the fiber $U(1)$.
The fiber direction is parametrized by $y=\psi\pm\varphi$.
Note that $L$ has no $\varphi$-dependence for $\theta=0,\pi$.
The zweibein of $S^2$ is given by the $i=1,2$ components of the left-invariant 1-form, 
$-iL^{-1}dL=\mu e^i\sigma_i/2$ \cite{Salam:1981xd}.
It takes the form
\begin{align}
&e^1=\frac{1}{\mu}(\pm\sin\varphi d\theta+\sin\theta\cos\varphi d\varphi), \n
&e^2=\frac{1}{\mu}(-\cos\varphi d\theta \pm \sin\theta\sin\varphi d\varphi).
\end{align}
This zweibein gives the standard metric of $S^2$ with the radius $1/\mu$:
\begin{align}
ds^2=\frac{1}{\mu^2}(d\theta^2+\sin^2\theta\varphi^2).
\label{metric of S^2}
\end{align}
Making a replacement $\partial_y \rightarrow -iq$ in (\ref{Killing vector})
leads to the angular momentum operator in the presence of a monopole with
magnetic charge $q$ at the origin \cite{Wu}:
\begin{eqnarray}
&&L_1^{(q)}=i(\sin\varphi\partial_{\theta}+
\cot\theta\cos\varphi\partial_{\varphi})-
q\frac{1\mp \cos\theta}{\sin\theta}\cos\varphi, \nonumber\\
&&L_2^{(q)}=i(-\cos\varphi\partial_{\theta}+
\cot\theta\sin\varphi\partial_{\varphi})-
q\frac{1\mp \cos\theta}{\sin\theta}\sin\varphi, \nonumber\\
&&L_3^{(q)}=-i\partial_{\varphi}\mp q ,
\label{monopole angular momentum}
\end{eqnarray}
where $q$ is quantized as 
$q=0, \pm \frac{1}{2}, \pm 1, \pm \frac{3}{2},\cdots$,
because $y$ is a periodic variable with the period $4\pi$.
These operators act on the local sections on $S^2$ and satisfy the $SU(2)$
algebra $[L_i^{(q)},L_j^{(q)}]=i\epsilon_{ijk}L_k^{(q)} $. 
Note that when $q=0$, these operators are reduced to the ordinary
angular momentum operators (\ref{angular momentum operator}) on $S^2$ (or $R^3$),
which generate the isometry group of $S^2$, $SU(2)$.
The $SU(2)$ acting on $g$ from left survives as the isometry of $S^2$.
Note that in 2+1 SYM on $R\times S^2$ the isometry of $S^2$ is included
in the $SU(2|4)$ symmetry as a subgroup.

\subsection{Dimensional reductions}
We dimensionally reduce the higher dimensional theories 
to the lower dimensional theories \cite{Kim-Klose-Plefka,Lin-Maldacena,IIST,IIST2}. 
We start with ${\cal N}=4$ SYM on $R\times S^3$:
\begin{align}
S_{R\times S^3}=\frac{1}{g_{R\times S^3}^2}\int_{R\times S^3} \frac{1}{2}\mbox{Tr}(F\wedge \ast F).
\label{YM on S^3}
\end{align}
We put $A=X_iE^i$ (note that we have ignored $A_t$). Then, the curvature 2-form is given by 
\begin{align}
F &=dA-iA\wedge A \nonumber\\
&=\partial_tX_i dt\wedge E^i+ i\mu {\cal L}_iX_jE^i\wedge E^j+X_i dE^i-iX_iX_j E^i\wedge E^j \nonumber\\
&=\partial_tX_i dt\wedge E^i
+\frac{1}{2}\epsilon_{ijk}\left(\mu X_k +i\epsilon_{klm}(\mu {\cal L}_lX_m
-\frac{1}{2}[X_l,X_m])\right)E^i \wedge E^j.
\label{curvature 2-form}
\end{align}
By using (\ref{curvature 2-form}), we rewrite (\ref{YM on S^3}) as
\begin{align}
S_{R\times S^3}=\frac{1}{g_{R\times S^3}^2}\int dt\frac{d\Omega_3}{(\mu/2)^3}
\mbox{Tr}\left(\frac{1}{2}(\partial_tX_i)^2
-\frac{1}{2}\left(\mu X_i +i\epsilon_{ijk}(\mu {\cal L}_jX_k
-\frac{1}{2}[X_j,X_k])\right)^2 \right).
\label{YM on S^3 2}
\end{align}
By dropping the $y$-derivatives in (\ref{YM on S^3 2}), we obtain 2+1 SYM on $R\times S^2$:
\begin{align}
S_{R\times S^2}
=\frac{1}{g_{R\times S^2}^2}\int dt\frac{d\Omega_2}{\mu^2} \mbox{Tr}\left(
\frac{1}{2}(\partial_tX_i)^2
 -\frac{1}{2}\left(\mu X_i+i\epsilon_{ijk}(\mu L^{(0)}_jX_k-\frac{1}{2}[X_j,X_k])\right)^2 \right),
\label{YM-higgs on S^2}
\end{align}
where $g_{R\times S^2}^2=\frac{\mu g_{R\times S^3}^2}{4\pi}$.
Thus we obtain 2+1 SYM on $R\times S^2$ from ${\cal N}=4$ SYM on $R\times S^3$ by dimensionally reducing 
the fiber direction of $S^3$ viewed as a $U(1)$ bundle over $S^2$.
One of two $SU(2)$'s that are the isometry of $S^3$ survives as 
the isometry of $S^2$. 
Correspondingly,  the superconformal symmetry, $SU(2,2|4)$,  
reduces to the $SU(2|4)$ symmetry.
It is convenient for us to
rewrite (\ref{YM-higgs on S^2}) using the gauge field and a Higgs field on $S^2$.
We decompose $X_i$ into the components tangential and horizontal to $S^2$ \cite{Maldacena:2002rb}:
\begin{align}
\vec{X}=\chi \vec{e}_r + a_1\vec{e}_{\varphi}
-a_2\vec{e}_{\theta},
\label{Y vector}
\end{align}
where 
$a_1$ and $a_2$ are the gauge field on $S^2$ and 
$\chi$ is the Higgs field on $S^2$. 
Substituting (\ref{Y vector}) into (\ref{YM-higgs on S^2}) leads to
\begin{align}
S_{R\times S^2}
=\frac{1}{g_{R\times S^2}^2}\int dt\frac{d\Omega_2}{\mu^2} \mbox{Tr}\left(\frac{1}{2}(\partial_t a_{a'})^2
+\frac{1}{2}(\partial_t\chi)^2-\frac{1}{2}(f_{12}-\mu\chi)^2-
\frac{1}{2}(D_{a'}\chi)^2 \right),
\label{YM-higgs on S^2 2}
\end{align}
where $a'$ run from 1 to 2.
Dropping all the derivatives in (\ref{YM-higgs on S^2}), we obtain
\begin{align}
S_{PW}
=\frac{1}{g_{PW}^2}\int \frac{dt}{\mu^2} \mbox{Tr}\left(
\frac{1}{2}(\partial_tX_i)^2
 -\frac{1}{2}\left(\mu X_i-\frac{i}{2}\epsilon_{ijk}[X_j,X_k]\right)^2 \right),
\label{matrix model}
\end{align}
where $g_{PW}^2=\frac{g_{R\times S^2}^2}{4\pi}$. Thus PWMM is obtained from 2+1 SYM on $R\times S^2$ by
a dimensional reduction. 
In this reduction, the $SU(2|4)$ symmetry is preserved.

\subsection{Vacua}
While ${\cal N}=4$ SYM on $R\times S^3$ possesses the unique vacuum, 
2+1 SYM on $R\times S^2$ and PWMM possess many nontrivial vacua \cite{Berenstein:2002jq,Lin-Maldacena}.
Let us see how those vacua are described. 
First, the vacuum configurations 
of (\ref{YM-higgs on S^2 2}) with the gauge group $U(M)$ are determined by 
\begin{align}
&f_{12}-\mu\chi=0, \nonumber\\
&D_{a'}\chi=0.
\label{eom for YM-higgs on S^2 2}
\end{align}
In the gauge in which $\chi$ is diagonal, (\ref{eom for YM-higgs on S^2 2}) is solved as
\begin{align}
&\hat{a}_1=0, \nonumber\\
&\hat{a}_2=-\frac{\cos\theta\mp 1}{\sin\theta}\hat{\chi}, \nonumber\\
&\hat{\chi}=\mu\mbox{diag}(\cdots,\underbrace{q_{s-1},\cdots,q_{s-1}}_{N_{s-1}},
\underbrace{q_{s},\cdots,q_{s}}_{N_{s}},\underbrace{q_{s+1},\cdots,q_{s+1}}_{N_{s+1}},\cdots),
\label{vacuum of YM-higgs on S^2 2}
\end{align}
where the gauge field takes the configurations of Dirac's monopoles, so that 
$q_s$ must be half-integers due to Dirac's quantization condition. Note also that $\sum_sN_s=M$.
Thus the vacua of 2+1 SYM on $R\times S^2$ are classified by the monopole charges $q_s$ and their degeneracies $N_s$.
The vacua preserve the $SU(2|4)$ symmetry.
(\ref{eom for YM-higgs on S^2 2}) is rewritten in terms of  the notation in (\ref{YM-higgs on S^2}) as
\begin{align}
\mu X_i+i\mu\epsilon_{ijk}L_j^{(0)}X_k-\frac{i}{2}\epsilon_{ijk}[X_j,X_k]=0,
\end{align}
which is equivalent to
\begin{align}
[\mu L_i^{(0)}-X_i,\mu L_j^{(0)}-X_j]=i\mu \epsilon_{ijk}(\mu L_k^{(0)}-X_k),
\end{align}
and (\ref{vacuum of YM-higgs on S^2 2}) is rewritten as 
\begin{align}
\mu L_i^{(0)}-\hat{X}_i=\mu \mbox{diag}(\cdots,\underbrace{L_i^{(q_{s-1})},\cdots,L_i^{(q_{s-1})}}_{N_{s-1}},
\underbrace{L_i^{(q_{s})},\cdots,L_i^{(q_{s})}}_{N_{s}},\underbrace{L_i^{(q_{s+1})},\cdots,L_i^{(q_{s+1})}}_{N_{s+1}},
\cdots),
\label{monopole background}
\end{align}
where $\vec{\hat{X}}=\hat{\chi}\vec{e}_r+\hat{a}_1\vec{e}_{\varphi}-\hat{a}_2\vec{e}_{\theta}$.

Next, the vacuum configurations of (\ref{matrix model}) with the gauge group $U(\hat{M})$ 
are determined by
\begin{align}
[X_i,X_j]=-i\mu\epsilon_{ijk}X_k.
\label{eom for matrix model}
\end{align}
(\ref{eom for matrix model}) is solved as 
\begin{align}
\hat{X}_i=-\mu L_i,
\label{vacuum of matrix model}
\end{align}
where $L_i$ are the representation matrices of the $SU(2)$ generators
which are in general
reducible, and are decomposed into irreducible representations:
\begin{align}
L_i=
 \begin{pmatrix}
  \rotatebox[origin=tl]{-35}
  {$\cdots \;\;\;
  \overbrace{\rotatebox[origin=c]{35}{$L_{i}^{[j_{s-1}]}$} \;
  \cdots \;
  \rotatebox[origin=c]{35}{$L_{i}^{[j_{s-1}]}$}}^{\rotatebox{35}{$N_{s-1}$}}
  \;\;\;
  \overbrace{\rotatebox[origin=c]{35}{$L_i^{[j_{s}]}$} \;
  \cdots \;
  \rotatebox[origin=c]{35}{$L_i^{[j_{s}]}$}}^{\rotatebox{35}{$N_{s}$}}
  \;\;\;
  \overbrace{\rotatebox[origin=c]{35}{$L_{i}^{[j_{s+1}]}$} \;
  \cdots \;
  \rotatebox[origin=c]{35}{$L_{i}^{[j_{s+1}]}$}}^{\rotatebox{35}{$N_{s+1}$}}
  \;\;\;\cdots $}
 \end{pmatrix},
 \label{matrix background}
\end{align}
where $L_i^{[j]}$ are the spin $j$ representation matrices of $SU(2)$ and 
$\sum_s N_s(2j_s+1)=\hat{M}$. The vacua of the matrix model are classified by the $SU(2)$ representations $[j_s]$
and their degeneracies $N_s$. (\ref{matrix background}) represents concentric fuzzy spheres with different radii.
The vacua preserve the $SU(2|4)$ symmetry.

\subsection{Higher dimensional theories from lower dimensional theories}
In what follows, we obtain the higher dimensional theories from the lower dimensional theories.
First, we recall the relationship between the theory around (\ref{monopole background}) of
$2+1$ SYM on $R\times S^2$ and the theory around (\ref{vacuum of matrix model}) of PWMM, 
which was shown in \cite{Maldacena:2002rb} for the trivial vacuum of 2+1 SYM on $R\times S^2$ and 
in \cite{ISTT} for generic vacua.
We introduce an ultraviolet cutoff $N_0$ and put
\begin{align}
& 2q_s=2j_s+1-N_0\\
&\frac{4\pi}{g_{R\times S^2}^2}=\frac{N_0}{g_{PW}^2}.
\end{align}
Then, the theory around (\ref{monopole background}) is equivalent to the theory around (\ref{vacuum of matrix model})
in the limit in which $N_0\rightarrow \infty$
with $q_{s}$ and $g_{R\times S^2}$ fixed.
The equivalence is proved as follows.
We decompose the fields into the background corresponding to (\ref{monopole background})
and the fluctuation as $X_i^{(s,t)} \rightarrow \hat{X}_i^{(s,t)}+X_i^{(s,t)}$, where $(s,t)$
label the (off-diagonal) blocks. Note that $X_i^{(s,t)}$ is 
an $N_s\times N_t$ matrix. Then, (\ref{YM-higgs on S^2}) is expanded around
(\ref{monopole background}) as 
\begin{align}
S_{R\times S^2}&=\frac{1}{g_{R\times S^2}^2}\int dt\frac{d\Omega_2}{\mu^2} \frac{1}{2}\sum_{s,t}
\mbox{tr}\left[\partial_tX_i^{(s,t)}\partial_tX_i^{(t,s)}\right. \nonumber\\
&\qquad\qquad\qquad\qquad-\left(\mu X_i^{(s,t)}+i\mu\epsilon_{ijk}L^{(q_{st})}_jX_k^{(s,t)}
-\frac{i}{2}\epsilon_{ijk}[X_j,X_k]^{(s,t)}\right) \nonumber\\
&\left. \qquad\qquad\qquad\quad\qquad
\times\left(\mu X_i^{(t,s)}+i\mu\epsilon_{ilm}L^{(q_{ts})}_lX_m^{(t,s)}
-\frac{i}{2}\epsilon_{ilm}[X_l,X_m]^{(t,s)}\right)\right],
\label{YM-higgs on S^2 expanded around vacuum}
\end{align}
where 
\begin{align}
q_{st}=q_s-q_t.
\end{align}
We make a harmonic expansion of (\ref{YM-higgs on S^2 expanded around vacuum})
by expanding the fluctuation in terms of the monopole vector spherical harmonics $\tilde{Y}_{Jmqi}^{\rho}$:
\begin{align}
X_i^{(s,t)}=\sum_{\rho=0,\pm 1}\sum_{\tilde{Q}\geq |q_{st}|}
\sum_{m=-Q}^Q x_{Jm\rho}^{(s,t)}\tilde{Y}_{Jmqi}^{\rho},
\label{mode expansion of X}
\end{align}
where $\rho$ stands for the polarization,
$Q=J+\delta_{\rho 1}$ and $\tilde{Q}=J+\delta_{\rho -1}$.
The properties of the monopole spherical harmonics are analyzed and summarized in \cite{ISTT,IIST2,IIOST} and
references therein.
Substituting (\ref{mode expansion of X}) into (\ref{YM-higgs on S^2 expanded around vacuum})
yields
\begin{align}
S_{R\times S^2}&=\frac{4\pi}{g_{R\times S^2}^2}\int \frac{dt}{\mu^2}\mbox{tr}\left[
\frac{1}{2}\sum_{s,t}x_{Jm\rho}^{(s,t)\dagger}(\partial_t^2-\mu^2\rho^2 (J+1)^2)x_{Jm\rho}^{(s,t)} \right.\nonumber\\
&+i\mu\sum_{s,t,u}\rho_1(J_1+1){\cal E}_{J_1m_1q_{st}\rho_1\;J_2m_2q_{tu}\rho_2\;J_3m_3q_{us}\rho_3}
x_{J_1m_1\rho_1}^{(s,t)}x_{J_2m_2\rho_2}^{(t,u)}x_{J_3m_3\rho_3}^{(u,s)} \nonumber\\
&+\frac{1}{2}\sum_{s,t,u,v}(-1)^{m-q_{su}+1}
{\cal E}_{J-mq_{us}\rho\;J_1m_1q_{st}\rho_1\;J_2m_2q_{tu}\rho_2}
{\cal E}_{Jmq_{su}\rho\;J_3m_3q_{uv}\rho_3\;J_4m_4q_{vs}\rho_4} \nonumber\\
&\left.\qquad\qquad\qquad \times
x_{J_1m_1\rho_1}^{(s,t)}x_{J_2m_2\rho_2}^{(t,u)}x_{J_3m_3\rho_3}^{(u,v)}x_{J_4m_4\rho_4}^{(v,s)}
\right],
\label{mode expansion of YM-higgs on S^2}
\end{align}
where ${\cal E}_{J_1m_1q_{st}\rho_1\;J_2m_2q_{tu}\rho_2\;J_3m_3q_{us}\rho_3}$ is defined by
\begin{align}
&{\cal E}_{
J_1m_1q_1\rho_1
J_2m_2q_2\rho_2 
J_3m_3q_3\rho_3}  
=\int \frac{d\Omega_2}{4\pi}
\epsilon_{ijk}
\tilde{Y}^{\rho_1}_{J_1m_1q_1i}
\tilde{Y}^{\rho_2}_{J_2m_2q_2j}
\tilde{Y}^{\rho_3}_{J_3m_3q_3k}  \nonumber\\
&=
\sqrt{6(2J_1+1)(2J_1+2\rho_1^2+1)
       (2J_2+1)(2J_2+2\rho_2^2+1)
       (2J_3+1)(2J_3+2\rho_3^2+1)} \nonumber\\
&\;\;\;
\times (-1)^{-\frac{\rho_1+\rho_2+\rho_3+1}{2}}
\left\{
\begin{array}{ccc}
Q_1 & \tilde{Q}_1 & 1 \\
Q_2 & \tilde{Q}_2 & 1 \\
Q_3 & \tilde{Q}_3 & 1 
\end{array}
\right\}
\left(
\begin{array}{ccc}
Q_1 & Q_2 & Q_3 \\
m_1 & m_2 & m_3 
\end{array}
\right)
\left(
\begin{array}{ccc}
\tilde{Q}_1 & \tilde{Q}_2 & \tilde{Q}_3 \\
q_1         & q_2        & q_3 
\end{array}
\right),
\end{align}
and we have used the equality
\begin{align}
i\epsilon_{ijk}L_j^{(q)}\tilde{Y}^{\rho}_{Jmqk}
+\tilde{Y}^{\rho}_{Jmqi}
=\rho(J+1)\tilde{Y}^{\rho}_{Jmqi}.
\end{align}

Similarly, decomposing the matrices into the background given by (\ref{vacuum of matrix model})
and the fluctuation as $X_i \rightarrow \hat{X}_i+X_i$ leads to the theory around 
(\ref{vacuum of matrix model}):
\begin{align}
S_{PW}&=\frac{N_0}{g_{PW}^2}\int \frac{dt}{\mu^2} \frac{1}{2}\sum_{s,t}
\mbox{tr}\left[\partial_tX_i^{(s,t)}\partial_tX_i^{(t,s)}-\left(\mu X_i^{(s,t)}+i\mu\epsilon_{ijk}L_j\circ X_k^{(s,t)}
-\frac{i}{2}\epsilon_{ijk}[X_j,X_k]^{(s,t)}\right) \right.\nonumber\\
&\left. \qquad\qquad\qquad\qquad\qquad
\times\left(\mu X_i^{(t,s)}+i\mu\epsilon_{ilm}L_l\circ X_m^{(t,s)}
-\frac{i}{2}\epsilon_{ilm}[X_l,X_m]^{(t,s)}\right)\right],
\label{matrix model expanded around vacuum}
\end{align}
where $L_i\circ$ is defined by
\begin{align}
L_i\circ X_j^{(s,t)}=L_i^{[j_s]}X_j^{(s,t)}
-X_j^{(s,t)}L_i^{[j_t]}.
\end{align}
The gauge symmetry of the above theory is expressed as
\begin{align}
\delta X_i^{(s,t)}=i\mu L_i\circ \alpha^{(s,t)}-i[X_i,\alpha]^{(s,t)}.
\label{gauge symmetry}
\end{align}
We make a harmonic expansion of (\ref{matrix model expanded around vacuum})
by expanding the fluctuation in terms of the fuzzy vector spherical harmonics $\hat{Y}_{Jm(j_sj_t)i}^{\rho}$ 
defined in appendix A as
\begin{align}
X_i^{(s,t)}=\sum_{\rho=0,\pm 1}\sum_{\tilde{Q}\geq |j_s-j_t|}^{j_s+j_t}
\sum_{m=-Q}^Q x_{Jm\rho}^{(s,t)}\otimes \hat{Y}_{Jm(j_sj_t)i}^{\rho}.
\label{mode expansion of X 2}
\end{align}
$x_{Jm\rho}^{(s,t)}$ in (\ref{mode expansion of X 2}) is an $N_s\times N_t$ matrix.
Since $j_s+j_t=N_0+q_s+q_t-1$, $N_0$ plays the role of the 
ultraviolet cutoff. Note also that $j_s-j_t=q_s-q_t=q_{st}$.
Substituting (\ref{mode expansion of X 2}) into (\ref{matrix model expanded around vacuum}) yields
\begin{align}
S_{PW}&=\frac{N_0}{g_{PW}^2}\int \frac{dt}{\mu^2}\mbox{tr}\left[
\frac{1}{2}\sum_{s,t}x_{Jm\rho}^{(s,t)\dagger}(\partial_t^2-\rho^2 (J+1)^2)x_{Jm\rho}^{(s,t)} \right.\nonumber\\
&+i\mu\sum_{s,t,u}\rho_1(J_1+1)\hat{{\cal E}}_{J_1m_1(j_sj_t)\rho_1\;J_2m_2(j_tj_u)\rho_2\;J_3m_3(j_uj_s)\rho_3}
x_{J_1m_1\rho_1}^{(s,t)}x_{J_2m_2\rho_2}^{(t,u)}x_{J_3m_3\rho_3}^{(u,s)} \nonumber\\
&+\frac{1}{2}\sum_{s,t,u,v}(-1)^{m-q_{su}+1}
\hat{{\cal E}}_{J-m(j_uj_s)\rho\;J_1m_1(j_sj_t)\rho_1\;J_2m_2(j_tj_u)\rho_2}
\hat{{\cal E}}_{Jm(j_sj_u)\rho\;J_3m_3(j_uj_v)\rho_3\;J_4m_4(j_vj_s)\rho_4} \nonumber\\
&\left.\qquad\qquad\qquad \times
x_{J_1m_1\rho_1}^{(s,t)}x_{J_2m_2\rho_2}^{(t,u)}x_{J_3m_3\rho_3}^{(u,v)}x_{J_4m_4\rho_4}^{(v,s)}
\right],
\label{mode expansion of matrix model}
\end{align}
where $\hat{{\cal E}}_{J_1m_1(jj')\rho_1J_2m_2(j'j'')\rho_2J_3m_3(j''j)\rho_3}$
is defined in (\ref{cE})
and we have used 
(\ref{equalities for vector and spinor fuzzy spherical harmonics}).
In the $N_0\rightarrow \infty$ limit, the ultraviolet cutoff goes to infinity 
and 
\begin{align}
\hat{{\cal E}}_{J_1m_1(j_sj_t)\rho_1\;J_2m_2(j_tj_u)\rho_2\;J_3m_3(j_uj_s)\rho_3}
\rightarrow
{\cal E}_{J_1m_1q_{st}\rho_1\;J_2m_2q_{tu}\rho_2\;J_3m_3q_{us}\rho_3}
\label{reduction of cE}
\end{align}
because
the 6-j symbol behaves asymptotically for $R \gg 1$  as \cite{vmk}
\begin{equation}
\left\{
\begin{array}{ccc}
a   & b   & c \\
d+R & e+R & f+R \\
\end{array}
\right\}
\approx
\frac{(-1)^{a+b+c+2(d+e+f+R)}}{\sqrt{2R}}
\left(
\begin{array}{ccc}
a   & b   & c \\
e-f & f-d & d-e \\
\end{array}
\right).
\label{asymptotic form of 6j} 
\end{equation}
Namely, this limit corresponds to
the commutative (continuum) limit of the fuzzy spheres.
Hence, in the 
$N_0\rightarrow \infty$ limit with $g_{PW}^2/N_0=g_{R\times S^2}^2/(4\pi)$ and 
$q_s=j_s-\frac{N_0}{2}+\frac{1}{2}$ fixed, 
(\ref{mode expansion of matrix model}) agrees 
with (\ref{mode expansion of YM-higgs on S^2}). We have proven our statement.

This equivalence is classical in the following sense. 
The asymptotic formula (\ref{asymptotic form of 6j}) holds for $a,b,c \ll R$.
Namely, the reduction (\ref{reduction of cE})
is valid for $J_1,J_2,J_3 \ll N_0$.
Thus the equivalence is true at tree level. The loop effect may cause a deviation between
the two theories quantum mechanically, since in the loop $J_1,J_2,J_3$ can be ${\cal O}(N_0)$.
Part of this deviation should be attributed to the UV/IR mixing\footnote{
What we call the UV/IR mixing here is investigated as the noncommutative anomaly in \cite{Chu:2001xi,Panero:2006bx}}.
Suppose we restrict ourselves to the planar limit, 
in which $N_s\rightarrow\infty$ 
with $g_{R\times S^2}^2N_s$ fixed. Then, this restriction and
sixteen supersymmetries are probably sufficient to suppress the UV/IR mixing,
namely the noncommutativity in the continuum limit.
Furthermore, as we discuss later, they completely stabilize the vacua
of PWMM.
Thus, the equivalence should also hold at the quantum level.
Indeed, the gravity duals of 2+1 SYM on $R\times S^2$ and PWMM proposed in \cite{Lin-Maldacena} 
support this conjecture \cite{Ling,ISTT}.
In the next section, we give an evidence that the UV/IR mixing does not exist.

Next, we recall that the theory around a certain vacuum of 2+1 $U(M=N\times\infty)$ SYM on $R\times S^2$
with the orbifolding (periodicity) condition imposed is equivalent to $U(N)$ ${\cal N}=4$ SYM on $R\times S^3$, which was shown 
in \cite{ISTT} (See also \cite{IIST,IIST2}). 
This is an extension of the matrix T-duality to that on a nontrivial fiber
bundle, $S^3$ as a $U(1)$ bundle over $S^2$.
The vacuum of 2+1 SYM on $R\times S^2$
we take is given by (\ref{monopole background}) in the $N_0\rightarrow \infty$
limit with $s$ running from $-\infty$ to $\infty$,
$q_s=s/2$, $N_s=N$ and $\frac{4\pi}{\mu}g_{R\times S^2}^2=g_{R\times S^3}^2$. We decompose the fields on $S^2$ into the background and the fluctuation
\begin{align}
X_i \rightarrow \hat{X}_i+X_i 
\label{background and fluctuation}
\end{align}
and impose the periodicity (orbifolding) condition on the fluctuation
\begin{align}
X_i^{(s+1,t+1)}=X_i^{(s,t)}\equiv X_i^{(s-t)}.
\label{periodicity condition}
\end{align}
The fluctuations are gauge-transformed from
the patch I to the patch II as \cite{IIST} 
\begin{align}
X_i^{(s-t)}
=e^{-i(s-t)\varphi}X_i^{(s-t)}.
\label{gauge transformation of fluctuations}
\end{align}
We make the Fourier transformation for the fluctuations
on each patch to construct the gauge field
on the total space from:
\begin{align}
X_i(t,\theta,\varphi,\psi)&=\sum_{w}X_i^{(w)}(t,\theta,\varphi)
e^{-i\frac{1}{2}wy}.
\label{Fourier transformation}
\end{align}
We see from (\ref{gauge transformation of fluctuations}) that
the left-hand side of (\ref{Fourier transformation}) is indeed independent of the patches.
Using (\ref{Fourier transformation}), we obtain
\begin{align}
&L_i^{(q_{st})}X_j^{(s,t)}(\theta,\varphi)
=\frac{1}{4\pi}\int_0^{4\pi}dy {\cal L}_iX_j(\theta,\varphi,\psi)e^{i\frac{1}{2}(s-t)y}, \nonumber\\
&(X_i(\theta,\varphi)X_j(\theta,\varphi))^{(s,t)}
=\frac{1}{4\pi}\int_0^{4\pi}dy X_i(\theta,\varphi,\psi)X_j(\theta,\varphi,\psi)e^{i\frac{1}{2}(s-t)y},
\end{align}
and so on. Then, we see that (\ref{YM-higgs on S^2 expanded around vacuum}) equals 
$\sum_s \times$ (\ref{YM on S^3}).
We divide an overall factor $\sum_s$ to extract a single period and obtain $U(N)$ ${\cal N}=4$ SYM on $R\times S^3$.
Of course, we can verify this equivalence by seeing
the agreement of the harmonic 
expansions of the two theories.
We expand $X_i(t,\theta,\varphi,\psi)$ in terms of 
the vector spherical harmonics on $S^3$, 
$Y_{Jm\tilde{m}i}^{\rho}(\theta,\varphi,\psi)$ 
(See \cite{IIST2,IIOST,ISTT,ITT}):
\begin{align}
X_i(t,\theta,\varphi,\psi)
=\sum_{\rho=0,\pm 1}\sum_{J}
\sum_{m=-Q}^Q\sum_{\tilde{m}=-\tilde{Q}}^{\tilde{Q}}
x_{Jm\tilde{m}\rho}(t)Y_{Jm\tilde{m}i}^{\rho}(\theta,\varphi,\psi),
\end{align}
where $J$ run over all non-negative integers and half-integers.
$Q=J+\delta_{\rho 1}$ and $\tilde{Q}=J+\delta_{\rho -1}$
are the spins for the two $SU(2)$'s of the isometry of $S^3$.
Note that the $SU(2)$ whose spin is $\tilde{Q}$ is broken in (\ref{YM-higgs on S^2}).
By using the equality 
\begin{align}
\tilde{Y}_{Jmqi}^{\rho}(\theta,\varphi)
=e^{iqy}Y_{Jmqi}^{\rho}(\theta,\varphi,\psi),
\end{align}
we can easily show that the harmonic expansion of (\ref{YM on S^3 2})
agrees with (\ref{mode expansion of YM-higgs on S^2})$/\sum_{s}$
in the present set-up with the correspondence
\begin{align}
x_{Jm\rho}^{(s,t)} \leftrightarrow x_{Jm\frac{1}{2}(s-t)\rho}.
\end{align}
Namely, $\frac{1}{2}(s-t)$ is identified with $\tilde{m}$.

Combining the above two equivalences, we see that the theory 
around (\ref{vacuum of matrix model})
of PWMM in the $N_0\rightarrow\infty$ limit,
where $s$ runs from $-\infty$ to $\infty$, $2j_s+1=N_0+s$ and $N_s=N$,
is equivalent to
$U(N)$ ${\cal N}=4$ SYM on $R\times S^3$ if 
$g_{PW}^2/N_0$ is fixed to $\frac{g_{R\times S^3}^2\mu}{16\pi^2}$, 
the periodicity condition is imposed on the fluctuation and
the overall factor $\Sigma_s$ is divided. 

\subsection{Proposal for a nonperturbative definition of ${\cal N}=4$ SYM}
The relationship between ${\cal N}=4$ SYM on $R\times S^3$ and 2+1 SYM on $R\times S^2$
is again classical for the following reason, and 
so is the relationship between ${\cal N}=4$ SYM on $R\times S^3$
and PWMM. In order to construct a well-defined quantum theory,
we need to introduce an ultraviolet cutoff to the momentum of 
the fiber direction, which corresponds to $w$ in (\ref{Fourier transformation}).
Namely, we should consider finite-size matrices by making $s,t$ run from $-T/2$ to $T/2$ with $T$ an ultraviolet cutoff.
In this situation, however, the periodicity condition (\ref{periodicity condition}) is not compatible with
the gauge invariance. In order to resolve this problem, referring to the result for the modified reduced model
in the previous section, we discard the periodicity condition and take 
the limit in which $N_0\rightarrow\infty,\;T\rightarrow\infty,\;N\rightarrow\infty,\; T/N_0\rightarrow 0,\;
\frac{g_{PW}^2N}{N_0}=\frac{g_{R\times S^2}^2N}{4\pi}=\frac{g_{R\times S^3}^2N\mu}{16\pi^2}=\mbox{fixed}$. 
In this case, the $S^1$ in the previous section corresponds to
the fiber direction of $S^3$ viewed as a $U(1)$ bundle over $S^2$. 
Our theory is a one-dimensional massive theory, so the instability discussed in the last part of the previous section
is suppressed. Moreover, the $SU(2|4)$ symmetry preserved by the vacuum (\ref{vacuum of matrix model})
completely stabilizes the vacuum.
Indeed, the result in \cite{Dasgupta} ensures the perturbative stability, and it is easily seen from the result 
in \cite{Lin} that the nonperturbative instability 
via the tunneling to other vacua of PWMM caused by the instantons is suppressed in the $N\rightarrow\infty$ limit.
Thus, we do not need any quenching, and
we can respect the gauge symmetry and the $SU(2|4)$ symmetry, namely 
half of supersymmetries of ${\cal N}=4$ SYM on $R\times S^3$, simultaneously.
Indeed, (\ref{gauge symmetry}) should correspond to the gauge symmetry of 
${\cal N}=4$ SYM on $R\times S^3$.
The noncommutativity probably vanishes in the continuum limit as mentioned 
before.

To summarize, we propose a nonperturbative definition of ${\cal N}=4$ SYM on $R\times S^3$
as follows.
We consider the theory around (\ref{vacuum of matrix model}) of PWMM with 
\begin{align}
&-T/2 \leq s \leq T/2, \n
&2j_s+1=N_0+s, \n
&N_s=N. 
\label{background}
\end{align}
We take the limit in which
\begin{align}
&N_0\rightarrow\infty,\;\;\;T\rightarrow\infty,\;\;\;N\rightarrow\infty, \n
&\mbox{with} \;\; \frac{T}{N_0}\rightarrow 0 \;\;\mbox{and}\;\;
\frac{g_{PW}^2N}{N_0}=\frac{\mu}{16\pi^2}g_{R\times S^3}^2N=\mbox{fixed}. 
\label{limit}
\end{align}
Then, we obtain the 't Hooft (planar) limit of ${\cal N}=4$ SYM on $R\times S^3$.
The condition $T/N_0\rightarrow 0$ can be relaxed 
to $N_0-\frac{T}{2}\rightarrow\infty$ that should be required 
to obtain the continuum spheres.
For simplicity of the analysis, we adopt 
the stronger condition $T/N_0\rightarrow
0$ in this paper. The result should not depend on how to take the limit.
Our formulation preserves the gauge symmetry and the $SU(2|4)$ symmetry. It is, in particular, remarkable that
it preserves sixteen supersymmetries. We need to check the restoration of the superconformal symmetry $SU(2,2|4)$
to verify that our formulation does work well. The restoration of the superconformal symmetry should imply that
no UV/IR mixing occur. In the next section, we give some evidences for the restoration of the superconformal symmetry.

\section{Perturbative analysis}
\setcounter{equation}{0}
In this section, we perform a perturbative expansion of the theory around (\ref{vacuum of matrix model}) of PWMM. In the beginning, we do not assume
(\ref{background}) or (\ref{limit}).
We make a replacement $X_i\rightarrow \hat{X}_i+X_i$ in (\ref{action_of_PWMM}).
We adopt the Feynman-type gauge and add the following gauge fixing and Fadeev-Popov terms to the action: 
\begin{align}
\frac{1}{g_{PW}^2}\int \frac{dt}{\mu^2}\mbox{Tr}\left(-\frac{1}{2}(-\partial_tA_t+i\mu[L_i,X_i])^2
+i\bar{c}\partial_tD_tc+\mu\bar{c}[L_i,i\mu[L_i,c]-i[X_i,c]]\right).
\label{gauge fixing and F.-P. terms}
\end{align}
The resultant gauge-fixed action is written down in (\ref{gauge fixed action}) 
in appendix B.
The mode expansion of the fields is given in (\ref{mode_expansion_in_PWMM}), which of course includes (\ref{mode expansion of X 2}).
The harmonic expansion
of the gauge-fixed action is given in (\ref{harmonic_expansion_of_S_free}), 
(\ref{harmonic_expansion_of_S_gauge_int}) and 
(\ref{harmonic_expansion_of_S_matter_int}) which are a counterpart of (\ref{mode expansion of matrix model}). 
One can read off the propagators from (\ref{harmonic_expansion_of_S_free}) as in (\ref{propagators}) 
and the vertices from (\ref{harmonic_expansion_of_S_gauge_int}) 
and (\ref{harmonic_expansion_of_S_matter_int}).

\begin{figure}[tbp]
\begin{center}
\includegraphics[height=5cm, keepaspectratio, clip]{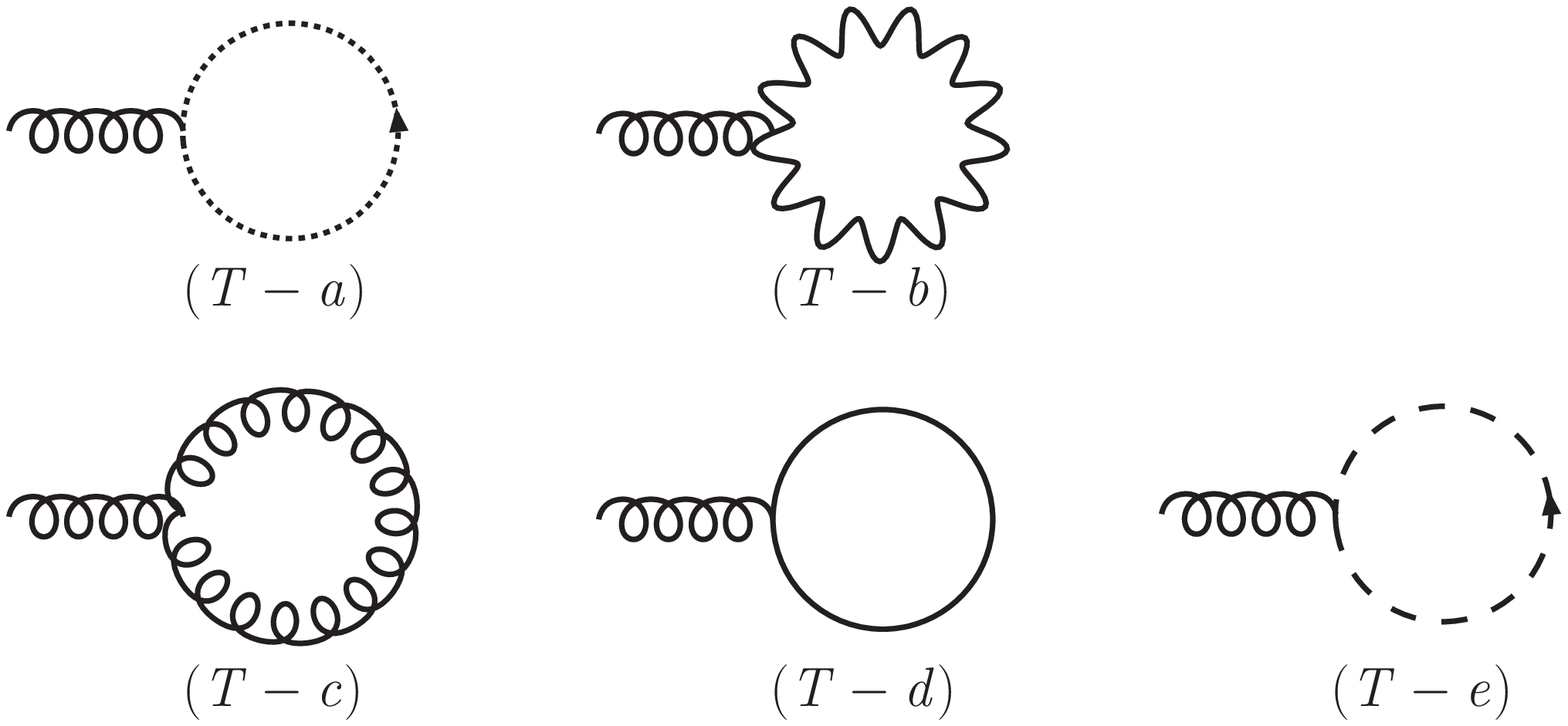}
\end{center}
\caption{Tadpole diagrams. The curly line
represents the propagator of $X_i$. The wavy line represents the propagator
of $A_t$. The dotted line represents the propagator of the ghost.
The solid line represents the propagator of $\Phi_{AB}$. The dashed line
represents the propagator of $\psi^A$.}
\end{figure}

First, we calculate the 1-loop contribution to the tadpoles. The only possibly nonzero contribution is
the truncated 1-point function for $x_{Jm\rho}^{(s,t)}(p)_{ij}$, where $i,j$ run from 1 to $N_s$ and $p$ is
dual to $t$ in the Fourier transformation.
This quantity takes the form
\begin{align}
2\pi\delta(p)\delta_{st}\delta_{ij}\delta_{\rho\: -1}\delta_{J0}\delta_{m0}\delta_{ij}\Upsilon^{(s)}.
\end{align}
There are five 1-loop diagrams for this 1-point function as shown in Fig. 2.
Note that all these diagrams are planar ones.
The diagrams $(T-a)$ and $(T-b)$ completely cancel each other.
Below we list the value of $\Upsilon^{(s)}$ for each of the remaining diagrams.
\begin{align}
&(T-c) 
=\frac{1}{2}g^2\sqrt{N_0}
\sum_{t,R}N_t(-1)^{R+j_s+j_t}\sqrt{R(R+1)(2R+1)}(2R+1) \nonumber\\
&\qquad\qquad\qquad\qquad\qquad \times\left(\frac{2R+3}{(R+1)^2}+\frac{2R-1}{R^2}+\frac{1}{R^{3/2}(R+1)^{3/2}}\right)
\sixj{1}{R}{R}{j_t}{j_s}{j_s}, \\
&(T-d) 
=12g^2\sqrt{N_0}
\sum_{t,R}N_t(-1)^{R+j_s+j_t}\sqrt{R(R+1)(2R+1)}
\sixj{1}{R}{R}{j_t}{j_s}{j_s}, \\
&(T-e) 
=-\frac{4}{3}\times (T-d),
\end{align}
where $g^2=g_{PW}^2\mu^2/N_0$.
The $6$-$j$ symbol in the above expressions can be written explicitly:
\begin{align}
\sixj{1}{R}{R}{j_t}{j_s}{j_s}
=(-1)^{j_t+j_s+R+1}\frac{1}{2}
\frac{(j_s-j_t)(j_s+j_t+1)+R(R+1)}
{\sqrt{j_s(j_s+1)(2j_s+1)R(R+1)(2R+1)}},
\label{explicit 6j}
\end{align}
By using (\ref{explicit 6j}) and $j_s=\frac{N_0}{2}+q_s-\frac{1}{2}$, we sum up the contributions to the tadpole:
\begin{align} 
&(T-c)+(T-d)+(T-e) \nonumber\\
&=g^2\sum_{t}N_t\sum_{R=|q_s-q_t|}^{N_0-1+q_s+q_t}F(R)
\left((q_s-q_t) 
+(R(R+1)+R\mbox{-independent terms}) \times {\cal O}\left(\frac{1}{N_0}\right)\right),
\end{align}
where 
\begin{align}
F(R)&=4-\frac{1}{2}(2R+1)\left(\frac{2R+3}{(R+1)^2}+\frac{2R-1}{R^2}+\frac{1}{R^{3/2}(R+1)^{3/2}}\right).
\label{F(R)}
\end{align}
The first term in (\ref{F(R)}) comes from $(T-d)$ and $(T-e)$ while the second term from $(T-c)$.
The asymptotic behavior of $F(R)$ for large $R$, 
\begin{align}
F(R) \sim \frac{3}{8R^4}+{\cal O}\left(\frac{1}{R^5}\right),
\end{align}
tells us that in the $N_0\rightarrow\infty$ limit 
\begin{align} 
(T-c)+(T-d)+(T-e)=g^2\sum_{t}N_t\sum_{R=|q_s-q_t|}^{\infty}F(R)(q_s-q_t)+{\cal O}\left(\frac{1}{N_0}\right).
\label{tadpole}
\end{align}
We find no $N_0$-dependent divergences.

We see that
\begin{align}
\sum_sN_s((T-c)+(T-d)+(T-e))=0.
\label{decoupling of U(1)}
\end{align}
This means that the vev of $\sum_{s,i}x_{Jm\rho}^{(s,s)}(p)_{ii}$ that corresponds to the one-point function for
the overall $U(1)$ field on $R\times S^2$ indeed vanishes.
This is consistent with the fact that it is a free field decoupled from the other fields when in the theory
around (\ref{monopole background}) one takes
the Feynman-like gauge, to which the gauge corresponding to (\ref{gauge fixing and F.-P. terms}) reduces
naively in the $N_0\rightarrow\infty$ limit.
One can easily verify from (\ref{F(R)}) that 
if there is no supersymmetry, 
the vev of the one-point function for the overall $U(1)$ field
does not vanish. Note that this happens 
even with the restriction to the planar limit
since all the tadpole diagrams in Fig. 2 are planar.
In \cite{CastroVillarreal:2004vh}, the same phenomenon was observed 
in a bosonic gauge theory on the fuzzy sphere 
in the continuum limit and
interpreted as the UV/IR mixing. 
On the other hand,
by shifting $\mu$ in (\ref{vacuum of matrix model}), one can always
cancel the vev of the one point function for the 
overall U(1) field and might obtain the commutative gauge theory.
However, in any case,
we cannot follow this prescription because it breaks supersymmetry.
Here we have obtained an evidence that in our case
the UV/IR mixing is avoided, that is, the noncommutativity vanishes
in the continuum limit,  in a way compatible with supersymmetry.

If we consider the theory around (\ref{vacuum of matrix model}) with (\ref{background}) and (\ref{limit})
that would realize ${N=4}$ SYM on $R\times S^3$, we find no $T$-dependent
divergences in (\ref{tadpole}). Furthermore, (\ref{tadpole}) vanishes for
fixed $s$ in the $T\rightarrow\infty$ due to the summation over $t$. 
In this case, the gauge corresponding to (\ref{gauge fixing and F.-P. terms}) reduces naively in the limit (\ref{limit})
to the Feynman gauge in ${\cal N}=4$ SYM on $R\times S^3$. The isometry of $S^3$,
$SO(4)=SU(2)\times SU(2)$, is manifest in ${\cal N}=4$ SYM on $R\times S^3$ with the Feynman gauge, and
all the tadpoles vanish due to this isometry.
The symmetry corresponding to one of the above two $SU(2)$'s already 
exists a priori in our theory, while the other does not.
Vanishing of (\ref{tadpole}) is a signal for the restoration of the $SO(4)$ symmetry in the continuum limit 
(\ref{limit}). 
If this restoration and the vanishing of the noncommutativity
is indeed the case, we obtain a commutative gauge theory 
with sixteen supersymmetries on $R\times S^3$.
This theory should be nothing but ${\cal N}=4$ SYM on $R\times S^3$ unless we perform any extra fine-tunings.
Thus we have found an evidence that the superconformal symmetry is restored
and our formalism does work well.


\begin{figure}[tbp]
\begin{center}
 \begin{minipage}{0.2\hsize}
  \begin{center}
   {\includegraphics[width=30mm]{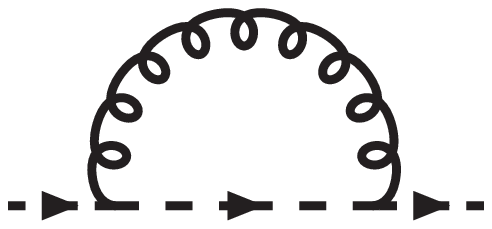}}
   $(F-a)$
   \label{fig:A1}
\end{center}

 \end{minipage}
 \begin{minipage}{0.2\hsize}
  \begin{center}
   {\includegraphics[width=30mm]{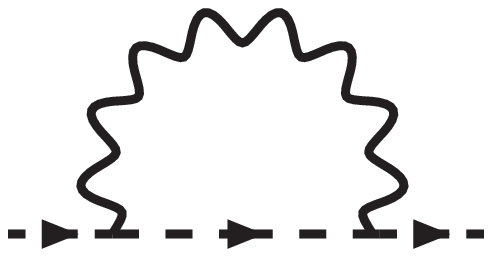}}
   $(F-b)$
   \label{fig:F2}
  \end{center}
 \end{minipage}
 \begin{minipage}{0.2\hsize}
  \begin{center}
   {\includegraphics[width=30mm]{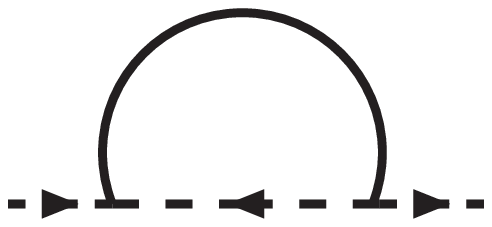}}
   $(F-c)$ 
   \label{fig:A6}
  \end{center}
 \end{minipage}
\end{center}
\caption{Diagrams for the one-loop self-energy of $\psi^A$. The curly line
represents the propagator of $X_i$. The wavy line represents the propagator
of $A_t$. 
The solid line represents the propagator of $\Phi_{AB}$. The dashed line
represents the propagator of $\psi^A$.}
\end{figure}

Next, we calculate the fermion self-energy
in the theory around (\ref{vacuum of matrix model}) with (\ref{background}) and (\ref{limit})
at the one-loop level,
and compare the result in ${\cal N}=4$ SYM on $R\times S^3$.
The fermion self-energy is given by the truncated two-point function $\langle \psi_{Jm\kappa}^{(s,t)A}(p)_{ij}\psi_{J'm'\kappa'A'}^{(s',t')\dagger}(p')_{kl}\rangle$, and 
this takes the form
\begin{align}
2\pi\delta(p-p')\delta_{ss'}\delta_{tt'}\delta^A_{A'}\delta_{il}\delta_{jk}
\delta_{JJ'}\delta_{mm'}\Omega^{(s,t)}_{J\kappa\kappa'}(p).
\end{align}
The diagrams which contribute to the fermion self-energy at the one-loop order
are shown in Fig. 3.
We list the value of $\Omega^{(s,t)}_{J\kappa\kappa'}(p)$ for each diagram in appendix C.

We set the external indices to specific values to calculate
the leading contribution in the continuum limit (\ref{limit}): 
$s=t=0$, $\kappa=\kappa'=1$ and $J=0$.
The divergent part of $\Omega_{011}^{(0,0)}(p)$ for each diagram 
is evaluated as
\begin{align}
&(F-a)_{\kappa=\kappa'=1,J=0}\n
&= -\frac{ig^2}{\mu^2}N
\sum_{u=-\frac{T}{2}}^{\frac{T}{2}}
\sum_{R_1=|\frac{u}{2}|}^{N_0-1+\frac{u}{2}}
\Biggl[
\frac{2R_1+3}{R_1+1} \cdot \frac{1}{l-(2R_1+\frac{7}{4})}
+\frac{2R_1-1}{R_1} \cdot \frac{1}{l+(2R_1+\frac{1}{4})}\n
&\quad
+\sqrt{\frac{R_1}{R_1+1}}
\cdot \frac{1}{l-\{R_1+\sqrt{R_1(R_1+1)}+\frac{3}{4}\}}
+\sqrt{\frac{R_1+1}{R_1}}\cdot
\frac{1}{l+\{R_1+\sqrt{R_1(R_1+1)}+\frac{1}{4}\}}
\Biggr] \n
&\sim -\frac{4ig^2}{\mu^2}N\left(-l-\frac{1}{4}\right)\ln T
-\frac{4ig^2}{\mu^2}N\left(-\frac{l}{2}+\frac{5}{8}\right)\ln T,
\n[3mm]
&(F-b)_{\kappa=\kappa'=1,J=0}\n
&=\frac{ig^2}{\mu^2}N
\sum_{u=-\frac{T}{2}}^{\frac{T}{2}}
\sum_{R_1=|\frac{u}{2}|}^{N_0-1+\frac{u}{2}} \n
&\quad \times \Biggl[
\sqrt{\frac{R_1+1}{R_1}}
\cdot \frac{1}{l-\{R_1+\sqrt{R_1(R_1+1)}+\frac{3}{4}\}}
+\sqrt{\frac{R_1}{R_1+1}}\cdot
\frac{1}{l+\{R_1+\sqrt{R_1(R_1+1)}+\frac{1}{4}\}}
\Biggr] \n
&\sim -\frac{4ig^2}{\mu^2}N\left(\frac{l}{2}+\frac{3}{8}\right)\ln T,
\n[3mm]
&(F-c)_{\kappa=\kappa'=1,J=0}\n
&=-\frac{12ig^2}{\mu^2}N
\sum_{u=-\frac{T}{2}}^{\frac{T}{2}}
\sum_{R_1=|\frac{u}{2}|}^{N_0-1+\frac{u}{2}}
\Biggl[
\frac{R_1+1}{2R_1+1}\cdot \frac{1}{l+(2R_1+\frac{5}{4})}
+\frac{R_1}{2R_1+1}\cdot \frac{1}{l-(2R_1+\frac{3}{4})}
\Biggr]\n
&\sim -\frac{12ig^2}{\mu^2}N\left(-l+\frac{3}{4}\right)\ln T,
\label{R=0 case}
\end{align}
where $l=p/\mu$.
Note that we find no $N_0$-dependent divergences in each diagram. This is consistent with the fact that (2+1)-dimensional gauge theory is super
renormalizable.
We find that the divergence in $T$ is logarithmic in each diagram.
This is again consistent with the fact that the fermion self-energy
has only the logarithmic divergence in four dimensions.

In ${\cal N}=4$ SYM on $R\times S^3$, 
the fermion self-energy is given by the two point function
$\langle \psi^{A}_{Jm\tilde{m}\kappa}(p)_{ij}
\psi^{\dagger}_{J'm'\tilde{m}'\kappa'A'}(p')_{kl} \rangle$,
where $J$ and $\tilde{J}$ are the spins for the two $SU(2)$'s of the isometry of $S^3$.
In the Feynman gauge to which the gauge corresponding to 
(\ref{gauge fixing and F.-P. terms}) reduces naively in the limit (\ref{limit}),
it takes the form
\begin{align}
2\pi\delta(p-p')\delta^A_{A'}\delta_{il}\delta_{jk}
\delta_{JJ'}\delta_{mm'}\delta_{\tilde{m}\tilde{m}'}\delta_{\kappa\kappa'}
\Omega^{S^3}_{J}(p).
\end{align}
By using the technique in \cite{ITT}, we evaluate each diagram in Fig. 3
in the Feynman
gauge. As in \cite{ITT}, we introduce naive cutoffs for the angular 
momenta and 
evaluate the divergent part of $\Omega^{S^3}_{J}(p)$ 
for each diagram as
\begin{align}
&(F-a)_{S^3}\n
&=\frac{4ig^2}{\mu^2} N 
\left[
\left\{ l+\frac{1}{3}\left(J+\frac{3}{4}\right)\right\} \ln (2\Lambda)
+\left\{ \frac{l}{2}-\frac{5}{6}\left(J+\frac{3}{4}\right)\right\}
\ln (2\Lambda)\right],\n
&(F-b)_{S^3}\n
&=\frac{4ig^2}{\mu^2} N 
\times \Bigl(-\frac{1}{2}\Bigr)
\left\{l+\left(J+\frac{3}{4}\right)\right\}\ln (2\Lambda),\n
&(F-c)_{S^3}\n
&=\frac{4ig^2}{\mu^2} N 
\times 3 \times
\left\{l-\left(J+\frac{3}{4}\right)\right\} \ln (2\Lambda).
\label{FS in S3}
\end{align}
The cutoffs for the angular momenta break
the gauge symmetry and supersymmetry. 
Nevertheless, the coefficient of 
the logarithmic divergent part for each diagram
has a universal meaning.
We find that (\ref{R=0 case}) completely agrees with 
the continuum case (\ref{FS in S3}) under the identification
$\Lambda=T$.
This fact provides an evidence that in the continuum limit (\ref{limit})
our theory reproduces  ${\cal N}=4$ SYM on $R\times S^3$.

Furthermore, 
we put $s=1$, $t=0$, $\kappa=1,\kappa'=-1$ and $J=\frac{1}{2}$.
For $\kappa\neq\kappa'$, the fermion self-energy
in ${\cal N}=4$ SYM on $R\times S^3$ in the Feynman gauge vanishes due to the $SO(4)$ symmetry,
while it does not vanish a priori in our theory
because the $SO(4)$ symmetry is not manifest in our theory. 
However, it turns out that there is no divergence in $\Omega_{\frac{1}{2}1\:-1}(p)$ for each diagram
and it vanishes due to the summation over the blocks 
(the summation over $u$ in (\ref{R=0 case})).
This is another evidence for the restoration of the $SO(4)$ symmetry, which implies the restoration of
the superconformal symmetry if the noncommutativity vanishes.

\section{Conclusion and discussion}
\setcounter{equation}{0}
In this paper, we proposed 
a nonperturbative definition of the 't Hooft limit of ${\cal N}=4$ SYM.
We realized ${\cal N}=4$ SYM on $R\times S^3$ as the theory around
a vacuum of PWMM. The size of matrices plays a role of the ultraviolet cutoff.
Our formulation preserves the gauge symmetry and the $SU(2|4)$ symmetry.
$SU(2|4)$ is a subgroup of $SU(2,2|4)$ which is 
the superconformal symmetry possessed by ${\cal N}=4$ SYM on $R\times S^3$. 
In particular, sixteen supersymmetries among 
thirty-two supersymmetries in 
${\cal N}=4$ SYM on $R\times S^3$ are preserved in our formulation.
We calculated the tadpoles and the fermion self-energy at the one-loop order. The results give
some evidences that 
the UV/IR mixing does not exist and 
the $SU(2,2|4)$ symmetry is restored in the continuum limit so that 
our formulation does work well.

We should collect more evidences for the restoration of 
the superconformal symmetry. Higher-loop calculations are needed.
Of course, the restoration should eventually be confirmed nonperturbatively.

The numerical simulation for our theory 
can be performed based on the method in \cite{Hanada,Anagnostopoulos,Catterall:2008yz}.
Unfortunately, the size of matrices available at present seems 
too small for the 
continuum limit (\ref{limit}). 
It is now possible to perform the numerical simulation for the theory
around (\ref{vacuum of matrix model}), for instance, with $s$ taking 
only 1 and $N_1=N$ and
to take the continuum limit that would realize 
the 't Hooft limit of 2+1 SYM on $R\times S^2$.
Then, we can compare the results of the numerical simulation with those
predicted by the gravity dual \cite{Lin-Maldacena} to
check whether the UV/IR mixing is avoided. 
Anyway, we believe that the numerical simulation for our theory will
be possible in the near future.
It is also desirable to develop 
an analytical (approximation) method that enables us to analyze our theory
at strong coupling.

By using the result in \cite{IIOST},
we can easily construct as a physical observable
the Wilson loop in our theory 
that corresponds to the ordinary Wilson loop
in ${\cal N}=4$ SYM on $R\times S^3$.
We can also consider the BPS Wilson loop \cite{Erickson,Drukker} 
by including the matter degrees of freedom in the loop.
It is important to calculate the vev of these Wilson loops in our theory
analytically and numerically in the strong coupling regime
and compare the results with the predictions of the gravity dual
\cite{Maldacena:1998im,Rey:1998ik,Alday:2007he}.
We also hope to find the integrable structure of ${\cal N}=4$ SYM at
strong coupling by analyzing our theory.

\section*{Acknowledgements}
We would like to thank S. Iso, H. Kawai, Y. Kitazawa, J. Nishimura, H. Suzuki
and K. Yoshida 
for discussions.
The work of G.I. and S.S. is supported in part by the JSPS Research Fellowship for Young
Scientists.
The work of A.T. is supported in part by Grant-in-Aid for Scientific
Research (No. 19540294) from the Ministry 
of Education, Culture, Sports, Science and Technology.

\appendix

\section{Fuzzy spherical harmonics \label{Fuzzy sherical harmonics}}
In this appendix, we summarize the properties of the fuzzy spherical harmonics analyzed and summarized in
\cite{ISTT} (see also \cite{Grosse:1995jt,Baez:1998he,Hoppe,deWit,Hoppe2,Dasgupta:2002hx}).
Let us consider $(2j+1)\times (2j'+1)$ rectangular complex matrices.
Such matrices are generally expressed as
\begin{align}
M=\sum_{r,r'}M_{rr'}|jr\rangle\langle j'r'|.
\end{align}
We can define linear maps $L_i\circ$, which map the set of 
$(2j+1)\times (2j'+1)$ rectangular complex matrices to itself, by their 
operation on the basis:
\begin{align}
L_i \circ |jr\rangle\langle j'r'| \equiv 
L_i^{[j]}|jr\rangle\langle j'r'|-
|jr\rangle\langle j'r'|L_i^{[j']},
\end{align}
where $L_i^{[j]}$ are the spin $j$ representation matrices of
the $SU(2)$ generators. $L_i \circ$ satisfy the 
$SU(2)$ algebra 
$[L_i\circ ,L_j\circ] = i\epsilon_{ijk}L_k \circ$.

We change the basis of the rectangular matrices 
from the above basis $\{|jr\rangle\langle j'r'|\}$
to the new basis which is called the fuzzy spherical harmonics:
\begin{align}
\hat{Y}_{Jm(jj')}=\sqrt{N_0}\sum_{r,r'}(-1)^{-j+r'}C^{Jm}_{jr\;j'-r'}
|jr\rangle\langle j'r'|,
\label{fuzzy spherical harmonics}
\end{align}
where $N_0$ is a positive constant, which is taken to be an integer as an ultraviolet cutoff in section 3.
For a fixed $J$, the fuzzy spherical harmonics also form the 
basis of the spin $J$ irreducible representation of $SU(2)$ which is
generated by $L_i \circ$
\begin{align}
(L_i \circ)^2\hat{Y}_{Jm(jj')}
&=J(J+1)\hat{Y}_{Jm(jj')},\nonumber\\
L_{\pm}\circ \hat{Y}_{Jm(jj')}
&=\sqrt{(J\mp m)(J \pm
 m+1)}\hat{Y}_{Jm\pm 1(jj')},\nonumber\\
L_3\circ \hat{Y}_{Jm(jj')}
&=m\hat{Y}_{Jm(jj')}.
\end{align}
The hermitian conjugates of the fuzzy spherical harmonics 
are evaluated as
\begin{align}
\left(\hat{Y}_{Jm(jj')}\right)^{\dagger}
=(-1)^{m-(j-j')}\hat{Y}_{J-m(j'j)}.
\end{align}
The fuzzy spherical harmonics satisfy the orthonormality
condition under the following normalized trace:
\begin{align}
\frac{1}{N_0}{\rm tr}\left\{
\left(\hat{Y}_{Jm(jj')}\right)^{\dagger}
\hat{Y}_{J'm'(jj')}\right\}=\delta_{JJ'}\delta_{mm'},
\end{align}
where $"{\rm tr}"$ stands for the trace over 
$(2j'+1)\times(2j'+1)$ matrices. 
The trace of three fuzzy spherical harmonics is given by
\begin{align}
\hat{C}^{J_1m_1(jj'')}_{J_2m_2(jj')J_3m_3(j'j'')}
&\equiv
\frac{1}{N_0}{\rm tr} \left\{
\left(\hat{Y}_{J_1m_1(jj'')}\right)^{\dagger}
\hat{Y}_{J_2m_2(jj')}\hat{Y}_{J_3m_3(j'j'')}
\right\} \nonumber\\
&=
(-1)^{J_1+j+j''}\sqrt{N_0(2J_2+1)(2J_3+1)}
C^{J_1m_1}_{J_2m_2J_3m_3}
\left\{
\begin{array}{ccc}
J_1 & J_2 & J_3 \\
j' & j'' & j  
\end{array}
\right\},
\end{align} 
where the last factor of the last line in 
the above equation is the $6$-$j$ symbol.

We also introduce the vector fuzzy spherical harmonics $\hat{Y}_{Jm(jj')i}^{\rho}$ and the spinor fuzzy
spherical harmonics $\hat{Y}_{Jm(jj')\alpha}^{\kappa}$, where $\rho$ takes -1,0,1 and $\kappa$ takes -1 and 1. They are defined in terms of the scalar spherical harmonics as
\begin{align}
&\hat{Y}_{Jm(jj')i}^{\rho}=i^\rho \sum_{n,p} V_{in}\cg{Qm}{\tilde{Q}p}{1n}\hat{Y}_{\tilde{Q}p(jj')}, \n
&\hat{Y}_{Jm(jj')\alpha}^{\kappa}=\sum_{p} \cg{Um}{\tilde{U}p}{\frac{1}{2}\alpha} \hat{Y}_{\tilde{U}p(jj')},
\end{align}
where $Q=J+\delta_{\rho1},\; \tilde{Q}=J+\delta_{\rho-1}$ and 
$U=J+\delta_{\kappa1},\; \tilde{U}=J+\delta_{\kappa-1}$.
The unitary matrix $V$ is given by
\begin{align}
V=\frac{1}{\sqrt{2}}
\begin{pmatrix}
-1 & 0 & 1 \\
-i & 0 & -i \\
0 & \sqrt{2} & 0
\end{pmatrix}.
\end{align}
The vector fuzzy spherical harmonics and the spinor fuzzy spherical harmonics
satisfy
\begin{align}
&L_i\circ \hat{Y}_{Jm(jj')i}^{\rho}
=\sqrt{J(J+1)}\delta_{\rho0}\hat{Y}_{Jm(jj')}, \n
&i\epsilon_{ijk}L_j\circ \hat{Y}_{Jm(jj')k}^{\rho}+\hat{Y}_{Jm(jj')i}^{\rho}
=\rho (J+1) \hat{Y}_{Jm(jj')i}^{\rho}, \n
&\left( (\sigma_i)_{\alpha\beta} L_i\circ +\frac{3}{4}\delta_{\alpha\beta}\right)\hat{Y}_{Jm(jj')\beta}^{\kappa}
=\kappa \left(J+\frac{3}{4}\right) \hat{Y}_{Jm(jj')\alpha}^{\kappa}.
\label{equalities for vector and spinor fuzzy spherical harmonics}
\end{align}
Their hermitian conjugate are
\begin{align}
&\left(\hat{Y}_{Jm(jj')i}^{\rho}\right)^\dagger
=(-1)^{m-(j-j')+1}\hat{Y}_{J-m(j'j)i}^{\rho}, \n
&\left(\hat{Y}_{Jm(jj')\alpha}^{\kappa}\right)^\dagger
=(-1)^{m-(j-j')+\kappa \alpha +1}\hat{Y}_{J-m(j'j)-\alpha}^{\kappa},
\end{align}
and they satisfy the following orthonormal relations:
\begin{align}
&\frac{1}{N_0}\tr\left\{
\left(\hat{Y}_{Jm(jj')i}^{\rho}\right)^\dagger
\hat{Y}_{J'm'(jj')i}^{\rho'}\right\}
=\delta_{JJ'}\delta_{mm'}\delta_{\rho\rho'}, \n
&\frac{1}{N_0}\tr\left\{
\left(\hat{Y}_{Jm(jj')\alpha}^{\kappa}\right)^\dagger
\hat{Y}_{J'm'(jj')\alpha}^{\kappa'}\right\}
=\delta_{JJ'}\delta_{mm'}\delta_{\kappa\kappa'}.
\end{align}
We can evaluate the trace of the three fuzzy spherical harmonics, including the vector harmonics and/or the spinor harmonics, as follows:
\begin{align}
\hat{\cD}^{Jm(j'j)}_{J_1m_1(j'j'')\rho_1\; J_2m_2(j''j)\rho_2}
&\equiv\frac{1}{N_0}\tr\left\{
\left(\hat{Y}_{Jm(j'j)}\right)^\dagger
\hat{Y}_{J_1m_1(j'j'')i}^{\rho_1}
\hat{Y}_{J_2m_2(j''j)i}^{\rho_2}
\right\} \n
&=\sqrt{3N_0(2J+1)(2J_1+1)(2J_1+2\rho_1^2+1)(2J_2+1)(2J_2+2\rho_2^2+1)}\n
& \qquad \times (-1)^{\frac{\rho_1+\rho_2}{2}+1+J+j+j'}
\ninej
{Q_1}{\tilde{Q}_1}{1}
{Q_2}{\tilde{Q}_2}{1}
{J}{J}{0}
\cg{Jm}{Q_1m_1}{Q_2m_2}
\sixj
{J}{\tilde{Q}_1}{\tilde{Q}_2}
{j''}{j}{j'}.
\label{cD}
\end{align}
\begin{align}
&\hat{{\cal E}}_{
J_1m_1(jj')\rho_1
J_2m_2(j'j'')\rho_2
J_3m_3(j''j)\rho_3}  \n
&\equiv\epsilon_{ijk}
\frac{1}{N_0} {\rm tr}
\left(
\hat{Y}^{\rho_1}_{J_1m_1(jj')i}
\hat{Y}^{\rho_2}_{J_2m_2(j'j'')j}
\hat{Y}^{\rho_3}_{J_3m_3(j''j)k}   
\right) \nonumber\\
&=
\sqrt{6N_0(2J_1+1)(2J_1+2\rho_1^2+1)
       (2J_2+1)(2J_2+2\rho_2^2+1)
       (2J_3+1)(2J_3+2\rho_3^2+1)} \nonumber\\
&\;\;\;
\times (-1)^{-\frac{\rho_1+\rho_2+\rho_3+1}{2}
             -\tilde{Q}_1-\tilde{Q}_2-\tilde{Q}_3
             +2j+2j'+2j''}
\ninej
{Q_1}{\tilde{Q}_1}{1} 
{Q_2}{\tilde{Q}_2}{1}
{Q_3}{\tilde{Q}_3}{1}
\left(
\begin{array}{ccc}
Q_1 & Q_2 & Q_3 \\
m_1 & m_2 & m_3 
\end{array}
\right)
\sixj
{\tilde{Q}_1}{\tilde{Q}_2}{\tilde{Q}_3}
{j''}{j}{j'}.
\label{cE}
\end{align}
\begin{align}
\hat{\cF}^{J_1m_1(j'j)\kappa_1}_{J_2m_2(j'j'')\kappa_2\; Jm(j''j)}
&\equiv
\frac{1}{N_0}\tr \left\{
\left(\hat{Y}_{J_1m_1(j'j)\alpha}^{\kappa_1}\right)^\dagger
\hat{Y}_{J_2m_2(j'j'')\alpha}^{\kappa_2}
\hat{Y}_{Jm(j''j)}
\right\} \n
&=\sqrt{2N_0(2\tilde{U}_1+1)(2J+1)^2(2J_2+1)(2J_2+2)}\n
&\qquad \times (-1)^{\tilde{U}_1+2J+j+j'}
\ninej
{U_1}{\tilde{U}_1}{\frac{1}{2}}
{U_2}{\tilde{U}_2}{\frac{1}{2}}
{J}{J}{0}
\cg{U_1m_1}{U_2m_2}{Jm}
\sixj
{\tilde{U}_1}{\tilde{U}_2}{J}
{j''}{j}{j'}.
\label{cF}
\end{align}
\begin{align}
\hat{\cG}^{J_1m_1(j'j)\kappa_1}_{J_2m_2(j'j'')\kappa_2\; Jm(j''j)\rho}
&\equiv
\frac{1}{N_0}\tr \left\{
\left(\hat{Y}_{J_1m_1(j'j)\alpha}^{\kappa_1}\right)^\dagger
\sigma_{\alpha\beta}^{i}
\hat{Y}_{J_2m_2(j'j'')\beta}^{\kappa_2}
\hat{Y}_{Jm(j''j)i}^{\rho}
\right\} \n
&=\sqrt{6N_0(2\tilde{U}_1+1)(2J_2+1)(2J_2+2)(2J+1)(2J+2\rho^2+1)}\n
&\qquad \times (-1)^{\frac{\rho}{2}+\tilde{U}_1+j+j'}
\ninej
{U_1}{\tilde{U}_1}{\frac{1}{2}}
{U_2}{\tilde{U}_2}{\frac{1}{2}}
{Q}{\tilde{Q}}{1}
\cg{U_1m_1}{U_2m_2}{Qm}
\sixj
{\tilde{U}_1}{\tilde{U}_2}{\tilde{Q}}
{j''}{j}{j'}.
\label{cG}
\end{align}

\section{Harmonic expansion
\label{Mode expansion of the fields and the action}}

\setcounter{equation}{0}
In this appendix, we make a harmonic expansion of the theory around 
(\ref{vacuum of matrix model}) of PWMM. This harmonic expansion enables us
to perform the perturbative calculation of the theory in section 4.
First, we make a replacement $X_i \rightarrow -\mu L_i +X_i$ in
(\ref{action_of_PWMM}) and add the gauge fixing and the Fadeev-Popov terms
(\ref{gauge fixing and F.-P. terms}).
The resultant action is
\begin{align}
S_{PW+gf+FP}
= S^{gauge}_{PW,free} + S^{gauge}_{PW,int}
+ S^{matter}_{PW,free} + S^{matter}_{PW,int},
\label{gauge fixed action}
\end{align}
where
\begin{align}
S^{gauge}_{PW,free}
&=\frac{1}{g_{PW}^2 \mu^2}\int dt \; \mbox{Tr}
\left(
\frac{1}{2}(\partial_t X_i)^2-\frac{\mu^2}{2}[L_i,A_t]^2
-\frac{1}{2}(\partial_t A_t)^2
\right. \nonumber\\
&\qquad\qquad\left.
-\frac{\mu^2}{2}(X_i+i\epsilon_{ijk}[L_j,X_k])^2
+\frac{\mu^2}{2}[L_i,X_i]^2
+i\bar{c}\partial_t^2c
+i\mu^2 \bar{c}[L_i,[L_i,c]]
\right),
\label{S_gauge_free}\displaybreak[0] \\
S^{gauge}_{PW,int}
&=\frac{1}{g_{PW}^2\mu^2}\int dt \; \mbox{Tr}
\biggl(
-i(\partial_t X_i)[A_t,X_i]-\mu[A_t,X_i][L_i,A_t]
-\frac{1}{2}[A_t,X_i]^2 \n
&\qquad\qquad
+i\mu\epsilon_{ijk}(X_i+i\epsilon_{ilm}[L_l,X_m])X_j X_k
+\frac{1}{2}\epsilon_{ijk}\epsilon_{ilm}X_j X_k X_l X_m \n
&\qquad \qquad
-i\mu [L_i,\bar{c}][c,X_i]
-\partial_t\bar{c}[A_t,c]
\biggr),
\label{S_gauge_int}\displaybreak[0]\\
S^{matter}_{PW,free}
&=\frac{1}{g_{PW}^2\mu^2}\int dt \; \mbox{Tr}
\left(
\frac{1}{2}\partial_t \Phi_{AB}\partial_t \Phi^{AB}
-\frac{\mu^2}{8}\Phi_{AB}\Phi^{AB}
+\frac{\mu^2}{2}[L_i,\Phi_{AB}][L_i,\Phi^{AB}]
\right. \nonumber\\
&\qquad\qquad\left.
-i\psi_A^\dagger \partial_t \psi^A
-\mu\psi_A^\dagger (\frac{3}{4}\psi^A+\sigma^i [L_i,\psi^A]\big)
\right), \label{S_matter_free}\displaybreak[0]\\
S^{matter}_{PW,int}
&=\frac{1}{g_{PW}^2\mu^2}\int dt \; \mbox{Tr}
\left(
-i(\partial_t \Phi_{AB})[A_t,\Phi^{AB}]
-\frac{1}{2}[A_t,\Phi_{AB}][A_t,\Phi^{AB}]
-\mu[L_i,\Phi_{AB}][X_i,\Phi^{AB}]
\right. \nonumber\\
&\qquad\qquad
+\frac{1}{2}[X_i,\Phi_{AB}][X_i,\Phi^{AB}] 
+\frac{1}{4}[\Phi_{AB},\Phi_{CD}][\Phi^{AB},\Phi^{CD}]
+\psi_A^\dagger [A_t,\psi^A]
\nonumber\\
&\qquad\qquad\left.
+\psi_A^\dagger \sigma^i [X_i,\psi^A]
+\psi_A^\dagger \sigma^2[\Phi^{AB},({\psi}_B^\dagger)^T]
-(\psi^A)^T \sigma^2 [\Phi_{AB},\psi^B]
\right).\label{S_matter_int}
\end{align}

We make a mode expansion of the $(s,t)$ blocks of the fields in terms of
the fuzzy spherical harmonics defined in appendix A:
\begin{align}
A_t^{(s,t)}
&=\sum_{J=|j_s-j_t|}^{j_s+j_t}\sum_{m=-J}^{J}B^{(s,t)}_{Jm}
\otimes\hat{Y}_{Jm(j_sj_t)}, 
\qquad
\Phi_{AB}^{(s,t)}
=\sum_{J=|j_s-j_t|}^{j_s+j_t}\sum_{m=-J}^{J}\phi^{(s,t)}_{AB,Jm}\otimes
\hat{Y}_{Jm(j_sj_t)},
\nonumber\displaybreak[0]\\
c^{(s,t)}
&=\sum_{J=|j_s-j_t|}^{j_s+j_t}\sum_{m=-J}^{J}c^{(s,t)}_{Jm}
\otimes\hat{Y}_{Jm(j_sj_t)}, 
\qquad
\bar{c}^{(s,t)}
=\sum_{J=|j_s-j_t|}^{j_s+j_t}\sum_{m=-J}^{J}\bar{c}^{(s,t)}_{Jm}\otimes
\hat{Y}_{Jm(j_sj_t)},
\nonumber\displaybreak[0]\\
\psi^{A(s,t)}
&=\sum_{\kappa=\pm1}\sum_{\tilde{U}=|j_s-j_t|}^{j_s+j_t}\sum_{m=-U}^{U}
\psi_{Jm\kappa}^{A(s,t)}\otimes\hat{Y}_{Jm(j_sj_t)}^\kappa \nonumber\\
&=\sum_{J=|j_s-j_t|}^{j_s+j_t}\sum_{m=-J-\frac{1}{2}}^{J+\frac{1}{2}}
\psi_{Jm1}^{A(s,t)}\otimes\hat{Y}_{Jm(j_sj_t)}^1
+\sum_{J=|j_s-j_t|-\frac{1}{2}}^{j_s+j_t-\frac{1}{2}}\sum_{m=-J}^{J}
\psi_{Jm-1}^{A(s,t)}\otimes\hat{Y}_{Jm(j_sj_t)}^{-1},
\nonumber\displaybreak[0]\\
\psi_A^{(t,s)\dagger}
&=\sum_{\kappa=\pm1}\sum_{\tilde{U}=|j_s-j_t|}^{j_s+j_t}\sum_{m=-U}^{U}
\psi_{A,Jm\kappa}^{(t,s)\dagger}\otimes\hat{Y}_{Jm(j_tj_s)}^{\kappa\dagger} \nonumber\\
&=\sum_{J=|j_s-j_t|}^{j_s+j_t}\sum_{m=-J-\frac{1}{2}}^{J+\frac{1}{2}}
\psi_{A,Jm1}^{(t,s)\dagger}\otimes\hat{Y}_{Jm(j_tj_s)}^{1\dagger}
+\sum_{J=|j_s-j_t|-\frac{1}{2}}^{j_s+j_t-\frac{1}{2}}\sum_{m=-J}^{J}
\psi_{A,Jm-1}^{(t,s)\dagger}\otimes\hat{Y}_{Jm(j_tj_s)}^{-1\dagger},
\nonumber\displaybreak[0]\\
X_i^{(s,t)}
&=\sum_{\rho=-1}^{1}\sum_{\tilde{Q}=|j_s-j_t|}^{j_s+j_t}\sum_{m=-Q}^{Q}
x_{Jm\rho}^{(s,t)}\otimes\hat{Y}{}_{Jm(j_sj_t)i}^\rho \nonumber\\
&=\sum_{J=|j_s-j_t|}^{j_s+j_t}\sum_{m=-J-1}^{J+1}
x_{Jm1}^{(s,t)}\otimes\hat{Y}{}_{Jm(j_sj_t)i}^1 
+\sum_{J=|j_s-j_t|}^{j_s+j_t}\sum_{m=-J}^{J}
x_{Jm0}^{(s,t)}\otimes\hat{Y}{}_{Jm(j_sj_t)i}^0 \nonumber\\
&\;\;\;\;\;+\sum_{J=|j_s-j_t|-1}^{j_s+j_t-1}\sum_{m=-J}^{J}
x_{Jm-1}^{(s,t)}\otimes\hat{Y}{}_{Jm(j_sj_t)i}^{-1}.
\label{mode_expansion_in_PWMM}
\end{align}
Note that the modes in the right-hand sides of the equations in
(\ref{mode_expansion_in_PWMM}) are $N_s\times N_t$ matrices.

By using the properties of the fuzzy spherical harmonics summarized in
appendix A, we rewrite the free part of (\ref{gauge fixed action})
in terms of the modes as follows:
\begin{align}
&S^{gauge}_{PW,free}+S^{matter}_{PW,free}\nonumber\\
&=\frac{1}{g^2} \int dt \; \mbox{tr}
\Biggl(
\frac{1}{2}(-1)^{m-(j_s-j_t)+1}x_{Jm\rho}^{(s,t)}
\bigl\{-\partial_t^2-\rho^2{\omega_J^x}^2-\mu^2\delta_{\rho0}J(J+1)\bigr\}
x_{J-m\rho}^{(t,s)} \n
&\qquad\qquad
+\frac{1}{2}(-1)^{m-(j_s-j_t)+1}B_{Jm}^{(s,t)}
\bigl\{-\partial_t^2-\mu^2 J(J+1)\bigr\}B_{J-m}^{(t,s)}\n
&\qquad\qquad
+i(-1)^{m-(j_s-j_t)}\bar{c}_{Jm}^{(s,t)}
\bigl\{\partial_t^2+\mu^2 J(J+1)\bigr\}c_{J-m}^{(t,s)}\n
&\qquad\qquad
+\frac{1}{4}(-1)^{m-(j_s-j_t)}\epsilon^{ABCD}\phi_{AB,Jm}^{(s,t)}
(-\partial_t^2-{\omega_J^x}^2)
\phi_{CD,J-m}^{(t,s)}
+\psi_{A,Jm\kappa}^{(s,t)\dagger}
(i\partial_t-\kappa\omega_J^\psi)\psi_{Jm\kappa}^{A(s,t)}
\Biggr),
\label{harmonic_expansion_of_S_free}
\end{align}
where
\begin{align}
&\frac{1}{g^2}\equiv\frac{N_0}{g_{PW}^2\mu^2},\;\;\omega_J^x\equiv\mu(J+1),
\nonumber \\
& \omega_J^\psi\equiv\mu(J+\frac{3}{4}),\;\;\omega_J^\phi\equiv\mu(J+\frac{1}{2}).
\end{align}
We can read off the propagators for the Fourier transforms of the fields
from (\ref{harmonic_expansion_of_S_free}) as
\begin{align}
&\langle x_{Jm\rho}^{(s,t)}(p)_{ij} x_{J'm'\rho'}^{(s',t')}(p')_{kl}\rangle
\n
&=
\begin{cases}
(-1)^{m-(j_s-j_t)+1}\delta_{JJ'}\delta_{m\,-m'}\delta_{\rho\rho'}
\delta_{st'}\delta_{ts'}\delta_{il}\delta_{jk}
2\pi\delta(p+p')\frac{ig^2}{p^2-{\omega_J^x}^2}
\;\;(\rho\neq0)\\
(-1)^{m-(j_s-j_t)+1}\delta_{JJ'}\delta_{m\,-m'}
\delta_{st'}\delta_{ts'}\delta_{il}\delta_{jk}
2\pi\delta(p+p')\frac{ig^2}{p^2-\mu^2J(J+1)}
\;\;(\rho=\rho'=0 )
\end{cases},
\displaybreak[0]\n
&\langle B_{Jm}^{(s,t)}(p)_{ij} B_{J'm'}^{(s',t')}(p')_{kl}\rangle
=
(-1)^{m-(j_s-j_t)+1}\delta_{JJ'}\delta_{m\,-m'}
\delta_{st'}\delta_{ts'}\delta_{il}\delta_{jk}
2\pi\delta(p+p')\frac{ig^2}{p^2-\mu^2 J(J+1)},
\displaybreak[0]\n
&\langle c_{Jm}^{(s,t)}(p)_{ij} \bar{c}_{J'm'}^{(s',t')}(p')_{kl}\rangle
=
(-1)^{m-(j_s-j_t)}\delta_{JJ'}\delta_{m\,-m'}
\delta_{st'}\delta_{ts'}\delta_{il}\delta_{jk}
2\pi\delta(p+p')\frac{g^2}{-p^2+\mu^2 J(J+1)},
\displaybreak[0]\n
&\langle \phi_{AB,Jm}^{(s,t)}(p)_{ij} \phi_{A'B',J'm'}^{(s',t')}(p')_{kl}
\rangle
=\frac{1}{2}\epsilon_{ABA'B'}
(-1)^{m-(j_s-j_t)}\delta_{JJ'}\delta_{m\,-m'}
\delta_{st'}\delta_{ts'}\delta_{il}\delta_{jk}
2\pi\delta(p+p')\frac{ig^2}{p^2-{\omega_J^\phi}^2},
\displaybreak[0]\n
&\langle \psi_{Jm\kappa}^{A(s,t)}(p)_{ij} \psi_{A',J'm'\kappa'}^{(s',t')\dagger}(p')_{kl}\rangle
=\delta_{JJ'}\delta_{mm'}\delta_{\kappa\kappa'}\delta^A_{A'}
\delta_{ss'}\delta_{tt'}\delta_{il}\delta_{jk}
2\pi\delta(p-p')\frac{ig^2(p+\kappa\omega_J^\psi)}{p^2-{\omega_J^\psi}^2}.
\label{propagators}
\end{align}

The gauge part of the interaction terms in (\ref{gauge fixed action}) is
rewritten as
\begin{align}
&S^{gauge}_{PW,int}\nonumber\\
&=\frac{1}{g^2} \int dt \; \mbox{tr}
\Big[
-i\hat{\mathcal{D}}_{J_1m_1(j_sj_t)\;J_2m_2(j_tj_u)\rho_2\;J_3m_3(j_uj_s)\rho_3}
B_{J_1m_1}^{(s,t)}
(\partial_t x_{J_2m_2\rho_2}^{(t,u)} x_{J_3m_3\rho_3}^{(u,s)}
-x_{J_2m_2\rho_2}^{(t,u)} \partial_t x_{J_3m_3\rho_3}^{(u,s)})
\nonumber \displaybreak[0]\\
&+\mu
(\sqrt{J_2(J_2+1)}\hat{\mathcal{D}}_{J_1m_1(j_sj_t)\;J_2m_2(j_tj_u)0\;J_3m_3(j_uj_s)\rho_3}
-\sqrt{J_1(J_1+1)}\hat{\mathcal{D}}_{J_2m_2(j_tj_u)\;J_3m_3(j_uj_s)\rho_3\;J_1m_1(j_sj_t)0})
\nonumber\\
&\quad\times
B_{J_1m_1}^{(s,t)} B_{J_2m_2}^{(t,u)} x_{J_3m_3\rho_3}^{(u,s)}
\nonumber\displaybreak[0]\\
&+(-1)^{m-(j_s-j_u)+1}
\nonumber\\
&\quad\times(\hat{\mathcal{D}}_{J_1m_1(j_sj_t)\;J_2m_2(j_tj_u)\rho_2\;J-m(j_uj_s)\rho}
\hat{\mathcal{D}}_{J_4m_4(j_vj_s)\;Jm(j_sj_t)\rho\;J_3m_3(j_uj_v)\rho_3}
B_{J_1m_1}^{(s,t)}  x_{J_2m_2\rho_2}^{(t,u)} x_{J_3m_3\rho_3}^{(u,v)}   B_{J_4m_4}^{(v,s)}
\nonumber\\
&\quad-\hat{\mathcal{D}}_{J_1m_1(j_sj_t)\;J_2m_2(j_tj_u)\rho_2\;J-m(j_uj_s)\rho}
\hat{\mathcal{D}}_{J_3m_3(j_uj_v)\;J_4m_4(j_vj_s)\rho_4\;Jm(j_sj_u)\rho}
B_{J_1m_1}^{(s,t)}  x_{J_2m_2\rho_2}^{(t,u)} B_{J_3m_3}^{(u,v)} x_{J_4m_4\rho_4}^{(v,s)})
\nonumber\displaybreak[0]\\
&+i\mu\rho_1(\rho_1+1)
\hat{\mathcal{E}}_{J_1m_1(j_sj_t)\rho_1\;J_2m_2(j_tj_u)\rho_2\;J_3m_3(j_uj_s)\rho_3}
x_{J_1m_1\rho_1}^{(s,t)} x_{J_2m_2\rho_2}^{(t,u)} x_{J_3m_3\rho_3}^{(u,s)}
\nonumber\displaybreak[0]\\
&+\frac{1}{2}(-1)^{m-(j_s-j_u)+1}
\hat{\mathcal{E}}_{J-m(j_uj_s)\rho\;J_1m_1(j_sj_t)\rho_1\;J_2m_2(j_tj_u)\rho_2}
\hat{\mathcal{E}}_{Jm(j_sj_u)\rho\;J_3m_3(j_uj_v)\rho_3\;J_4m_4(j_vj_s)\rho_4}
\nonumber\\
&\quad\times
x_{J_1m_1\rho_1}^{(s,t)} x_{J_2m_2\rho_2}^{(t,u)}
x_{J_3m_3\rho_3}^{(u,v)} x_{J_4m_4\rho_4}^{(v,s)}
\nonumber\displaybreak[0]\\
&-i\mu\sqrt{J_3(J_3+1)}\;
(\hat{\mathcal{D}}_{J_2m_2(j_sj_t)\;J_3m_3(j_tj_u)0\;J_1m_1(j_uj_s)\rho_1}
c_{J_2m_2}^{(s,t)} \bar{c}_{J_3m_3}^{(t,u)} x_{J_1m_1\rho_1}^{(u,s)}
\nonumber\\
&\quad-\hat{\mathcal{D}}_{J_2m_2(j_sj_t)\;J_1m_1(j_tj_u)\rho_1\;J_3m_3(j_uj_s)0}
c_{J_2m_2}^{(s,t)} x_{J_1m_1\rho_1}^{(t,u)} \bar{c}_{J_3m_3}^{(u,s)})
\nonumber\displaybreak[0]\\
&+\hat{\mathcal{C}}_{J_1m_1(j_sj_t)\;J_2m_2(j_tj_u)\;J_3m_3(j_uj_s)}
B_{J_1m_1}^{(s,t)}
(\partial_t \bar{c}_{J_2m_2}^{(t,u)} c_{J_3m_3}^{(u,s)}
+c_{J_2m_2}^{(t,u)}\partial_t \bar{c}_{J_3m_3}^{(u,s)})
\Big].
\label{harmonic_expansion_of_S_gauge_int}
\end{align}
The matter part of the interaction terms in (\ref{gauge fixed action}) is
rewritten as
{\small
\begin{align}
&S^{matter}_{PW,int}\nonumber\\
&=\frac{1}{g^2} \int dt \; \mbox{tr}
\bigg[\frac{i}{2}\epsilon^{ABCD}
\hat{\mathcal{C}}_{J_1m_1(j_sj_t)\;J_2m_2(j_tj_u)\;J_3m_3(j_uj_s)}
B_{J_1m_1}^{(s,t)}
(\partial_t \phi_{AB,J_2m_2}^{(t,u)} \phi_{CD,J_3m_3}^{(u,s)}
-\phi_{AB,J_2m_2}^{(t,u)}\partial_t \phi_{CD,J_3m_3}^{(u,s)})
\nonumber \displaybreak[0]\\
&+\frac{1}{2}\epsilon^{ABCD}
\hat{\mathcal{C}}^{Jm(j_sj_u)}_{J_1m_1(j_sj_t)\;J_2m_2(j_tj_u)}
\hat{\mathcal{C}}_{Jm(j_sj_u)\;J_3m_3(j_uj_v)\;J_4m_4(j_vj_s)}
\nonumber\\
&\quad\times
(B_{J_1m_1}^{(s,t)} B_{J_2m_2}^{(t,u)} \phi_{AB,J_3m_3}^{(u,v)} \phi_{CD,J_4m_4}^{(v,s)}
-B_{J_1m_1}^{(s,t)} \phi_{AB,J_2m_2}^{(t,u)} B_{J_3m_3}^{(u,v)} \phi_{CD,J_4m_4}^{(v,s)})
\nonumber\displaybreak[0]\\
&-\frac{\mu}{2}\epsilon^{ABCD}
(\sqrt{J_2(J_2+1)}
\hat{\mathcal{D}}_{J_1m_1(j_sj_t)\;J_2m_2(j_tj_u)0\;J_3m_3(j_uj_s)\rho_3}
\nonumber\\
&\quad-\sqrt{J_1(J_1+1)}
\hat{\mathcal{D}}_{J_2m_2(j_tj_u)\;J_3m_3(j_uj_s)\rho_3\;J_1m_1(j_sj_t)0})
 \phi_{AB,J_1m_1}^{(s,t)} \phi_{CD,J_2m_2}^{(t,u)} x_{J_3m_3\rho_3}^{(u,s)}
\nonumber\displaybreak[0]\\
&+\frac{1}{2}\epsilon^{ABCD}(-1)^{m-(j_s-j_u)+1}
\nonumber\\
&\quad\times
(\hat{\mathcal{D}}_{J_4m_4(j_vj_s)\;Jm(j_sj_u)\rho\;J_3m_3(j_uj_v)\rho_3}
\hat{\mathcal{D}}_{J_2m_2(j_tj_u)\;J-m(j_uj_s)\rho\;J_1m_1(j_sj_t)\rho_1}
 x_{J_1m_1\rho_1}^{(s,t)} \phi_{AB,J_2m_2}^{(t,u)}
 x_{J_3m_3\rho_3}^{(u,v)} \phi_{CD,J_4m_4}^{(v,s)}
\nonumber\\
&\quad
-\hat{\mathcal{D}}_{J_4m_4(j_uj_v)\;J_3m_3(j_vj_s)\rho_3\;Jm(j_sj_u)\rho}
\hat{\mathcal{D}}_{J_2m_2(j_tj_u)\;J-m(j_uj_s)\rho\;J_1m_1(j_sj_t)\rho_1}
 x_{J_1m_1\rho_1}^{(s,t)} \phi_{AB,J_2m_2}^{(t,u)}
 \phi_{CD,J_4m_4}^{(u,v)} x_{J_3m_3\rho_3}^{(v,s)})
\nonumber\displaybreak[0]\\
&+\frac{1}{8}\epsilon^{ABEF}\epsilon^{CDGH}
\hat{\mathcal{C}}^{Jm(j_sj_u)}_{J_1m_1(j_sj_t)\;J_2m_2(j_tj_u)}
\hat{\mathcal{C}}_{Jm(j_sj_u)\;J_3m_3(j_uj_v)\;J_4m_4(j_vj_s)}
\nonumber\\
&\quad\times
(\phi_{AB,J_1m_1}^{(s,t)} \phi_{CD,J_2m_2}^{(t,u)} \phi_{EF,J_3m_3}^{(u,v)} \phi_{GH,J_4m_4}^{(v,s)}
-\phi_{AB,J_1m_1}^{(s,t)} \phi_{EF,J_2m_2}^{(t,u)} \phi_{CD,J_3m_3}^{(u,v)} \phi_{GH,J_4m_4}^{(v,s)})
\nonumber\displaybreak[0]\\
&+\left((-1)^{m_3-(j_s-j_u)+\frac{\kappa_1-\kappa_2}{2}}
\hat{\mathcal{F}}^{J_2-m_2(j_tj_u)\kappa_2}_{J_1-m_1(j_tj_s)\kappa_1\;J_3m_3(j_sj_u)}
\psi_{A,J_1m_1\kappa_1}^{(s,t)\dagger} B_{J_3m_3}^{(s,u)} \psi_{J_2m_2\kappa_2}^{A(u,t)}
\right.\nonumber\\
&\quad\left.
-\hat{\mathcal{F}}^{J_1m_1(j_sj_t)\kappa_1}_{J_2m_2(j_sj_u)\kappa_2\;J_3m_3(j_uj_t)}
\psi_{A,J_1m_1\kappa_1}^{(s,t)\dagger} \psi_{J_2m_2\kappa_2}^{A(s,u)} B_{J_3m_3}^{(u,t)}
\right)\nonumber\displaybreak[0]\\
&-\left((-1)^{m_3-(j_s-j_u)+\frac{\kappa_1-\kappa_2}{2}}
\hat{\mathcal{G}}^{J_2-m_2(j_tj_u)\kappa_2}_{J_1-m_1(j_tj_s)\kappa_1\;J_3m_3(j_sj_u)\rho_3}
\psi_{A,J_1m_1\kappa_1}^{(s,t)\dagger} x_{J_3m_3\rho_3}^{(s,u)} \psi_{J_2m_2\kappa_2}^{A(u,t)}
\right.\nonumber\\
&\quad\left.
+\hat{\mathcal{G}}^{J_1m_1(j_sj_t)\kappa_1}_{J_2m_2(j_sj_u)\kappa_2\;J_3m_3(j_uj_t)\rho_3}
\psi_{A,J_1m_1\kappa_1}^{(s,t)\dagger} \psi_{J_2m_2\kappa_2}^{A(s,u)} x_{J_3m_3\rho_3}^{(u,t)}
\right)\nonumber\displaybreak[0]\\
&-\frac{i}{2}\epsilon^{ABCD}
\left(
(-1)^{m_1-(j_s-j_t)-\frac{\kappa_1}{2}}
\hat{\mathcal{F}}^{J_2m_2(j_tj_u)\kappa_2}_{J_1-m_1(j_tj_s)\kappa_1\;J_3m_3(j_sj_u)}
\psi_{A,J_1m_1\kappa_1}^{(s,t)\dagger} \phi_{CD,J_3m_3}^{(s,u)} \psi_{B,J_2m_2\kappa_2}^{(t,u)\dagger}
\right.\nonumber\\
&\quad\left.
+(-1)^{m_2-(j_u-j_s)-\frac{\kappa_2}{2}}
\hat{\mathcal{F}}^{J_1m_1(j_sj_t)\kappa_1}_{J_2-m_2(j_sj_u)\kappa_2\;J_3m_3(j_uj_t)}
\psi_{A,J_1m_1\kappa_1}^{(s,t)\dagger} \psi_{B,J_2m_2\kappa_2}^{(u,s)\dagger}
\phi_{CD,J_3m_3}^{(u,t)}
\right)\nonumber\displaybreak[0]\\
&-i\left((-1)^{m_2-(j_u-j_s)-\frac{\kappa_2}{2}}
\hat{\mathcal{F}}^{J_2-m_2(j_sj_u)\kappa_2}_{J_1m_1(j_sj_t)\kappa_1\;J_3m_3(j_tj_u)}
\psi_{J_1m_1\kappa_1}^{A,(s,t)} \phi_{AB,J_3m_3}^{(t,u)} \psi_{J_2m_2\kappa_2}^{B(u,s)}
\right.\nonumber\\
&\quad\left.
+(-1)^{m_1-(j_s-j_t)-\frac{\kappa_1}{2}}
\hat{\mathcal{F}}^{J_1-m_1(j_tj_s)\kappa_1}_{J_2m_2(j_tj_u)\kappa_2\;J_3m_3(j_uj_s)}
\psi_{J_1m_1\kappa_1}^{A(s,t)} \psi_{J_2m_2\kappa_2}^{B(t,u)} \phi_{AB,J_3m_3}^{(u,s)}
\right)
\bigg].
\label{harmonic_expansion_of_S_matter_int}
\end{align}
}

\section{Fermion self-energy}
\setcounter{equation}{0}
In this appendix, we list the value of $\Omega_{J\kappa\kappa'}^{(s,t)}(p)$ for
each diagram of the fermion self-energy in Fig. 3:
\begin{align}
 &(F-a) \n
 &=-\frac{3ig^2}{\mu^2}NN_0 \sqrt{(2\tilde{U}+1)(2\tilde{U'}+1)}
 \delta_{UU'}
 \sum_{uR_1R_2}\n
 &\Biggl[
 \frac{(2R_2+1)(2R_2+3)}{R_2+1}(2R_1+2)(2R_1+1)
 \frac{1}{l-(R_1+R_2+\frac{7}{4})}
 \ninej
 {U}{\tilde{U}}{\frac{1}{2}}
 {R_1+\thalf}{R_1}{\frac{1}{2}}
 {R_2+1}{R_2}{1}
 \ninej
 {U}{\tilde{U'}}{\frac{1}{2}}
 {R_1+\thalf}{R_1}{\frac{1}{2}}
 {R_2+1}{R_2}{1} \n
 &
 +\frac{(2R_2+1)(2R_2+3)}{R_2+1}2R_1(2R_1+1)\frac{1}{l+(R_1+R_2+\frac{5}{4})}
 \ninej
 {U}{\tilde{U}}{\frac{1}{2}}
 {R_1-\thalf}{R_1}{\frac{1}{2}}
 {R_2+1}{R_2}{1}
 \ninej
 {U}{\tilde{U'}}{\frac{1}{2}}
 {R_1-\thalf}{R_1}{\frac{1}{2}}
 {R_2+1}{R_2}{1} \n
 &
 +\frac{(2R_2-1)(2R_2+1)}{R_2}(2R_1+2)(2R_1+1)\frac{1}{l-(R_1+R_2+\frac{3}{4})}
 \ninej
 {U}{\tilde{U}}{\frac{1}{2}}
 {R_1+\thalf}{R_1}{\frac{1}{2}}
 {R_2-1}{R_2}{1}
 \ninej
 {U}{\tilde{U'}}{\frac{1}{2}}
 {R_1+\thalf}{R_1}{\frac{1}{2}}
 {R_2-1}{R_2}{1} \n
 &
 +\frac{(2R_2-1)(2R_2+1)}{R_2}2R_1(2R_1+1)\frac{1}{l+(R_1+R_2+\frac{1}{4})}
 \ninej
 {U}{\tilde{U}}{\frac{1}{2}}
 {R_1-\thalf}{R_1}{\frac{1}{2}}
 {R_2-1}{R_2}{1}
 \ninej
 {U}{\tilde{U'}}{\frac{1}{2}}
 {R_1-\thalf}{R_1}{\frac{1}{2}}
 {R_2-1}{R_2}{1} \n
&+\frac{(2R_2+1)^2}{\sqrt{R_2(R_2+1)}}
\times\Biggl(
\frac{(2R_1+2)(2R_1+1)}{l-(R_1+\sqrt{R_2(R_2+1)}+\frac{3}{4})}
\ninej
{U}{\tilde{U}}{\frac{1}{2}}
 {R_1+\thalf}{R_1}{\frac{1}{2}}
 {R_2}{R_2}{1}
\ninej
{U}{\tilde{U'}}{\frac{1}{2}}
 {R_1+\thalf}{R_1}{\frac{1}{2}}
 {R_2}{R_2}{1}\n
&+\frac{2R_1(2R_1+1)}{l+(R_1+\sqrt{R_2(R_2+1)}+\frac{1}{4})}
\ninej
{U}{\tilde{U}}{\frac{1}{2}}
 {R_1-\thalf}{R_1}{\frac{1}{2}}
 {R_2}{R_2}{1}
\ninej
{U}{\tilde{U'}}{\frac{1}{2}}
 {R_1-\thalf}{R_1}{\frac{1}{2}}
 {R_2}{R_2}{1}
\Biggr) \;
 \Biggr]\n
 &\qquad\qquad
 \times
 \Biggl[
 \sixj
 {\tilde{U}}{R_1}{R_2}
 {j_u}{j_s}{j_t}
 \sixj
 {\tilde{U'}}{R_1}{R_2}
 {j_u}{j_s}{j_t}
 +(-1)^{1-\frac{\kappa+\kappa'}{2}}
 \sixj
 {\tilde{U}}{R_1}{R_2}
 {j_u}{j_t}{j_s}
 \sixj
 {\tilde{U'}}{R_1}{R_2}
 {j_u}{j_t}{j_s}
 \Biggr],\\[3mm]
 &(F-b)\n
 &=\frac{ig^2}{\mu^2}NN_0 \sqrt{(2\tilde{U}+1)(2\tilde{U'}+1)}
 \delta_{UU'}
 \sum_{uR_1R_2}  \frac{(2R_2+1)^2}{\sqrt{R_2(R_2+1)}}\n
 &\times\Biggl[
 \frac{(2R_1+2)(2R_1+1)}{l-(R_1+\sqrt{R_2(R_2+1)}+\frac{3}{4})}
 \ninej
 {U}{\tilde{U}}{\frac{1}{2}}
 {R_1+\thalf}{R_1}{\frac{1}{2}}
 {R_2}{R_2}{0}
 \ninej
 {U}{\tilde{U'}}{\frac{1}{2}}
 {R_1+\thalf}{R_1}{\frac{1}{2}}
 {R_2}{R_2}{0}\n
 &+\frac{2R_1(2R_1+1)}{l+(R_1+\sqrt{R_2(R_2+1)}+\frac{1}{4})}
 \ninej
 {U}{\tilde{U}}{\frac{1}{2}}
 {R_1-\thalf}{R_1}{\frac{1}{2}}
 {R_2}{R_2}{0}
 \ninej
 {U}{\tilde{U'}}{\frac{1}{2}}
 {R_1-\thalf}{R_1}{\frac{1}{2}}
 {R_2}{R_2}{0}
 \Biggr]\n
 &\qquad \qquad \times
 \Biggl[
 \sixj
 {\tilde{U}}{R_1}{R_2}
 {j_u}{j_s}{j_t}
 \sixj
 {\tilde{U'}}{R_1}{R_2}
 {j_u}{j_s}{j_t}
 +(-1)^{1-\frac{\kappa+\kappa'}{2}}
 \sixj
 {\tilde{U}}{R_1}{R_2}
 {j_u}{j_t}{j_s}
 \sixj
 {\tilde{U'}}{R_1}{R_2}
 {j_u}{j_t}{j_s}
 \Biggr],\\[3mm]
 &(F-c)\n
 &=-\frac{12ig^2}{\mu^2}NN_0 \sqrt{(2\tilde{U}+1)(2\tilde{U'}+1)}
 \delta_{UU'}
 \sum_{uR_1R_2}  (2R_2+1) \n
 &\times\Biggl[
 (2R_1+2)(2R_1+1)\frac{1}{l+(R_1+R_2+\frac{5}{4})}
 \ninej
 {U}{\tilde{U}}{\frac{1}{2}}
 {R_1+\thalf}{R_1}{\frac{1}{2}}
 {R_2}{R_2}{0}
 \ninej
 {U}{\tilde{U'}}{\frac{1}{2}}
 {R_1+\thalf}{R_1}{\frac{1}{2}}
 {R_2}{R_2}{0}\n
 &+2R_1(2R_1+1)\frac{1}{l-(R_1+R_2+\frac{3}{4})}
 \ninej
 {U}{\tilde{U}}{\frac{1}{2}}
 {R_1-\thalf}{R_1}{\frac{1}{2}}
 {R_2}{R_2}{0}
 \ninej
 {U}{\tilde{U'}}{\frac{1}{2}}
 {R_1-\thalf}{R_1}{\frac{1}{2}}
 {R_2}{R_2}{0}
 \Biggr]\n
 &\qquad \qquad \times
 \Biggl[
 \sixj
 {\tilde{U}}{R_1}{R_2}
 {j_u}{j_s}{j_t}
 \sixj
 {\tilde{U'}}{R_1}{R_2}
 {j_u}{j_s}{j_t}
 +(-1)^{1-\frac{\kappa+\kappa'}{2}}
 \sixj
 {\tilde{U}}{R_1}{R_2}
 {j_u}{j_t}{j_s}
 \sixj
 {\tilde{U'}}{R_1}{R_2}
 {j_u}{j_t}{j_s}
 \Biggr],
\end{align}
where $l=p/\mu$.

\end{document}